\documentclass{emulateapj}

\usepackage{natbib}

\usepackage{amsmath,amssymb,amsfonts,bm}
\usepackage{hyperref}
\usepackage{calc}
\usepackage{color}
\usepackage{wrapfig}
\usepackage{url}
\usepackage{paralist}
\usepackage{array}
\usepackage{txfonts}
\slugcomment{}

\begin{document}

\title{Is Gravitational Lensing by Intercluster Filaments Always Negligible?}
\author{Martin Feix\altaffilmark{1}, Dong Xu\altaffilmark{2}, HuanYuan Shan\altaffilmark{3}, Benoit Famaey\altaffilmark{4}, Marceau Limousin\altaffilmark{2}, HongSheng Zhao\altaffilmark{1,3} and Andy Taylor\altaffilmark{5}}
\altaffiltext{1}{SUPA, School of Physics and Astronomy, University of St Andrews, North Haugh, St Andrews, Fife, KY16 9SS, United Kingdom}
\email{mf256@st-andrews.ac.uk (MF), dong@astro.ku.dk (DX)}
\altaffiltext{2}{Dark Cosmology Centre, Niels Bohr Institute, University of Copenhagen, Juliane Maries Vej 30, 2100, Copenhagen, Denmark}
\altaffiltext{3}{National Astronomical Observatories, Chinese Academy of Sciences, 20A Datun Road, Chaoyang District, 100012, Beijing, China}
\altaffiltext{4}{Institut d'Astronomie et d'Astrophysique, Universite Libre de Bruxelles, CP226, Bvd du Triomphe, B-1050, Bruxelles, Belgium}
\altaffiltext{5}{SUPA, Institute for Astronomy, University of Edinburgh, Royal Observatory, Blackford Hill, Edinburgh, EH9 3HJ, United Kingdom}

\begin{abstract}
Intercluster filaments negligibly contribute to the weak lensing signal in general relativity (GR), $\gamma_{N}\sim 10^{-4}-10^{-3}$.
In the context of relativistic modified Newtonian dynamics (MOND) introduced by Bekenstein, however, a single filament inclined by
$\approx 45^\circ$ from the line of sight can cause substantial distortion of background sources pointing towards
the filament's axis ($\kappa=\gamma=(1-A^{-1})/2\sim 0.01$); this is rigorous for infinitely long uniform filaments, but also qualitatively true
for short filaments ($\sim 30$Mpc), and even in regions where the projected matter density of the filament is equal to zero.
Since galaxies and galaxy clusters are generally embedded in filaments or are projected on such structures, 
this contribution complicates the interpretation of the weak lensing shear map in the context of MOND. While our analysis is of mainly theoretical interest providing order-of-magnitude estimates only, it seems safe to conclude that when modeling systems with anomalous weak lensing signals, e.g. the ``bullet cluster" of Clowe et al., the ``cosmic train wreck" of Abell 520 from Mahdavi et al., and the ``dark clusters" of Erben et al., {\it  filamentary structures might contribute} in a significant and likely complex fashion. On the other hand, {\it our predictions of a (conceptual) difference in the weak lensing signal could, in principle, be used to falsify MOND/TeVeS} and its variations.
\end{abstract}


\section{Introduction}
\label{intro}
Without resorting to cold dark matter (CDM), the modified Newtonian dynamics (MOND) paradigm \citep{Mond3, mondnew} is known to reproduce galaxy scaling relations like the Tully-Fisher relation \citep{tully}, the Faber-Jackson law \citep{faber} and the fundamental plane \citep{fundamental}) as well as the rotation curves of individual galaxies over five decades in mass \citep{spiral1,mondref1,mondref2,mondref3,mondref4,mondref5,escape}. In particular, the recent kinematic analysis of tidal dwarf galaxies by \cite{debris} is very hard to explain within the classical CDM framework while it is in accordance with MOND \citep{tidal1,tidal2}. In addition, observations of a tight correlation between the mass profiles of baryonic matter and dark matter in relatively isolated (field) galaxies at all radii \citep{insight2,insight} are most often interpreted as supporting MOND. Nevertheless, in rich clusters of galaxies, the MOND prescription is not enough to explain the observed discrepancy between visible and dynamical mass \citep{neutrinos2,tevesfit,asymmetric}. At very large radii, the discrepancy is about a factor of $2$, meaning that there should be as much dark matter (mainly in the central parts) as observed baryons in MOND clusters. One solution is that neutrinos have a mass at the limit of detection, i.e. $\sim2$ eV, which can solve the bulk of the problem of the missing mass in galaxy clusters, but other issues remain \citep{group}. These $2$ eV neutrinos have also been invoked to fit the angular power spectrum of the cosmic microwave background (CMB) in relativistic MOND \citep{tevesneutrinocosmo}, and are thus part of the only consistent MOND cosmology presented so far. In the following, we will refer to this model as the MOND hot dark matter ($\mu$HDM) cosmology \citep{tevesfit}\footnotemark\footnotetext[6]{Note, however, that one could also switch to sterile neutrinos with masses of a few eV \citep[e.g.][]{sterile,maltoni1,maltoni2} and that massive (sterile) neutrinos are not indispensable within certain covariant formulations of modified gravity, e.g. the $V\Lambda$ model, which can mimic the effects of neutrinos in clusters and cosmology as well as the behavior of a cosmological constant \citep{vector}.}.

On the other hand, strange features have recently been discovered in galaxy clusters, which are hard to explain, such as the ``dark matter core" devoid of galaxies at the center of the ``cosmic train wreck" cluster Abell 520 \citep{abell520} and others \citep{darkcluster,bullet}. Here, we consider the possibility that this kind of features could be due to the gravitational lensing effects generated by an intercluster filament in a universe based on tensor-vector-scalar gravity \citep[TeVeS;][]{teves}, one possible relativistic extension of MOND \citep[cf.][]{tv1,tv2,vector}. However, we are not performing a detailed lensing analysis of any particular cluster in the presence of filaments, but rather provide a proof of concept that the influence of filaments could be much less negligible in a MONDian universe than within the framework of general relativity (GR).

Filaments are among the most prominent large-scale structures of the universe. From simulations in $\Lambda$CDM cosmology, we know that almost every two neighboring clusters are connected by a straight filament with a length of approximately $20-30$ Mpc \citep{LCDMfilament}. For instance, the dynamics of field galaxies, which are generally embedded in such filaments, as well as their weak lensing properties are persistently influenced by this kind of
structures, generally encountering accelerations of about $0.01-0.1\times 10^{-10}$ m s$^{-2}$. 
Filaments also cover a fair fraction of the sky, much larger than the covering factor of galaxy
clusters. Thus, there is a good chance that filaments might be superimposed with other objects on a given line of sight,
hence affecting the analysis of observational data like, for example, weak lensing shear measurements. Such recent studies prompted us to investigate the possibility that, in the context of MOND, end-on filamentary structures could be responsible for creating anomalous features in reconstructions of weak lensing convergence maps such as the peculiar ``dark matter core" devoid of galaxies in Abell 520 \citep{abell520}.

Short straight filaments are structures which, at the best, are partially virialized in two directions perpendicular to their axis.
According to \cite{LCDMfilament}, a filament generally corresponds to an overdensity of about $10-30$, having a cigar-like shape. Furthermore, filamentary structures tend to have a low-density gradient along their axis and, in the perpendicular directions, they have a nearly uniform core which tapers to zero at larger radii, usually about $2-5$ times their core radius. Since filaments are typically much longer than their diameter, we shall approximately treat them as infinite uniform cylinders of radius $R_{f}=2.5$ $h^{-1}$ Mpc.

\begin{figure}
\centering
\includegraphics[trim=-20 0 0 0,width=\linewidth]{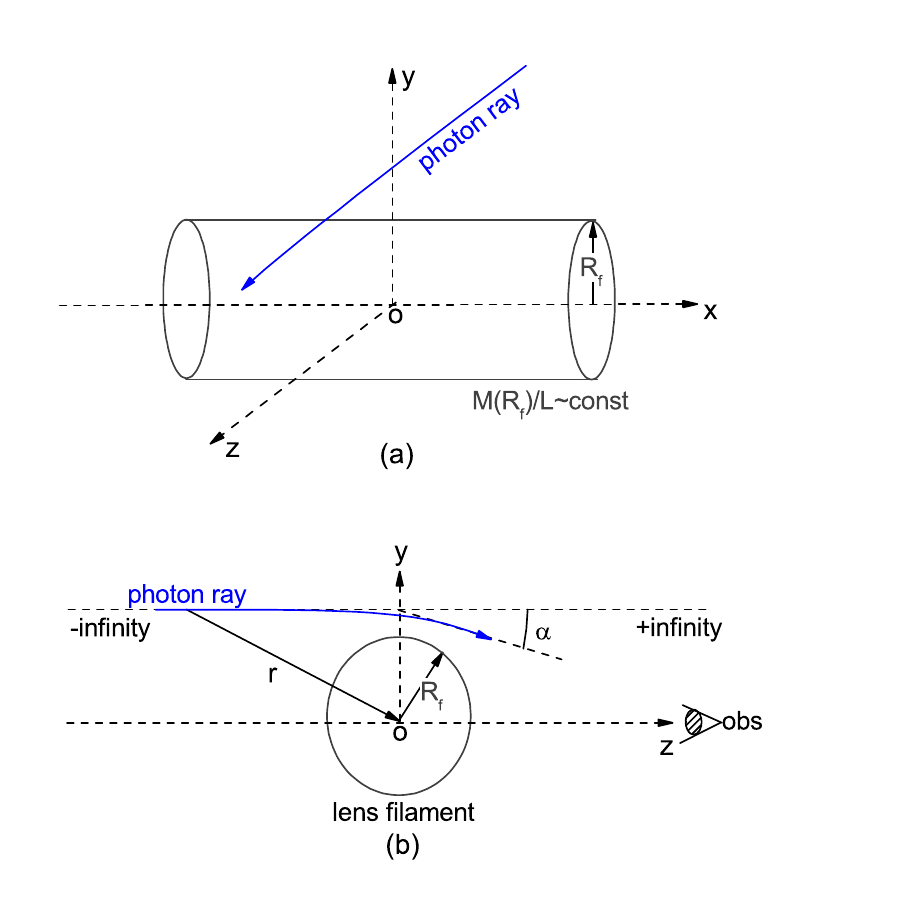}
\caption{Light deflection by an infinitely elongated cylinder of constant mass density; the unperturbed photon traveling
along the $z$-direction passes the filament at the distance $y$ (impact parameter) from the filament's axis and is
deflected by the angle $\hat\alpha$. The line density of the filament is assumed to be constant, $\lambda = M/L = \rho \pi R_f^2$, where $\rho$ is the volume density and $R_f$ is the cylinder's radius.}
\label{fig1}
\end{figure}

Lacking a MOND/TeVeS structure formation $N$-body simulation (with or without substantially massive neutrinos), we shall adopt the naive assumption that filamentary structures have roughly the same properties in MOND and in CDM, which will be justified in \S \ref{app}. Deriving expressions for the TeVeS deflection angle and setting up a cosmological background, we conclude that the order of magnitude of the TeVeS lensing signal caused by filaments is compatible with that of the previously mentioned observed anomalous systems. In addition, we find that there is fundamental difference between GR and MOND/TeVeS for cylindrically symmetric lens geometries (see Fig. \ref{fig1}); in contrast to GR, the framework of MOND/TeVeS allows us  to have image distortion and amplification effects where the projected matter density is equal to zero. As for a more realistic approach, we also consider a model where the filament has a fluctuating density profile perpendicular to its axis. Compared to the uniform model, we find that the lensing signal in this case is smaller, but still of the same order, taking into account that the filamentary structures may be inclined to the line of sight by rather small angles ($\theta\lesssim 20^{\circ}$). Finally, we demonstrate the impact of filaments onto the convergence map of other objects by considering superposition of such structures with a toy cluster along the line of sight. Again, our results show an additional contribution comparable to that of a single isolated filament.
\section{Modeling a Filamentary Lens}
\label{model}
We investigate the effect of gravitational lensing caused by a straight filament connecting two
galaxy clusters in both GR and TeVeS gravity, henceforth using units with $c=1$. As a first simple approach, we shall take the filament's matter density profile to equal an infinitely elongated and uniform cylinder which is illustrated in Figure \ref{fig1}.
The cylinder's line density,
\begin{equation}
\lambda = M/L = \rho \pi R_f^2,
\label{eq:linedef}
\end{equation}
is taken to be constant, where $M$ is the total mass, $L$ denotes the length along the symmetry axis, $R_f$ is the cylinder's radius, and $\rho$ is the volume density. A photon traveling perpendicular to the filament's axis will change its propagation direction when passing by the cylinder due to the local gravitational field which is assumed to be a weak perturbation to flat spacetime, i.e. all further calculations may be carried out within the non-relativistic approximation. In this case, it is well-known \citep{gl} that the photon's deflection angle can be expressed as
\begin{equation}
\vec{\hat{\alpha}} = 2\int\limits_{-\infty}^{\infty}{\vec\nabla}_{\bot}\Phi dl,
\label{eq:0}
\end{equation}
where $\Phi$ is the total gravitational potential, $\vec\nabla_{\bot}$ denotes the two-dimensional gradient operator perpendicular to light propagation and integration is performed along the unperturbed light path (Born's approximation).
In our example (see Fig. \ref{fig1}), the filament's axis is aligned with the $x$-axis, and light rays propagating along the $z$-direction are dragged into the $\pm y$-directions due to the symmetry of the resulting gravitational field. Keeping this configuration and introducing cylindrical coordinates, we may rewrite equation \eqref{eq:0} as
\begin{equation}
\hat\alpha(y) = 4y\int\limits_{y}^{\infty}\dfrac{\Phi^{'}}{\sqrt{r^{2}-y^{2}}}dr,
\label{eq:0a}
\end{equation}
where the prime denotes the derivative with respect to the cylindrical radial coordinate $r$, i.e. $A^{'}=dA/dr$. Note that even in the context of MOND/TeVeS, we may still assume that most of the light bending occurs within a small range around the lens compared to the distances between lens and source and observer and source, thus enabling us to fully adopt the GR lensing formalism.

In gravitational lensing, it is convenient to introduce the deflection potential $\Psi(\vec\theta)$ \citep{gl}:
\begin{equation}
\Psi(\vec\theta) = 2\frac{D_{ls}}{D_{s}D_{l}}\int\Phi(D_{l}\vec\theta,z)dz,
\label{eq:0b}
\end{equation}
where we have used $\vec\theta=\vec\xi/D_{l}$. Here $\vec\xi$ is the two-dimensional position vector in the lens plane, and $D_{s}$, $D_{l}$, and $D_{ls}$ are the (angular diameter) distances between source and observer, lens and observer, and lens and source, respectively. If a source is much smaller than the angular scale on which the lens properties change, the lens mapping can locally be linearized. Thus, the distortion of an image can be described by the Jacobian matrix
\begin{equation}
\mathcal{A}(\vec\theta) = \frac{\partial\vec\beta}{\partial\vec\theta} =
\begin{pmatrix}
1-\kappa-\gamma_{1} & -\gamma_{2}\\
-\gamma_{2} & 1-\kappa+\gamma_{1}
\end{pmatrix},
\label{eq:0c}
\end{equation}
where $\vec\beta$=$\vec\eta/D_{s}$ and $\vec\eta$ denotes the 2-dimensional position of the source. The convergence $\kappa$ is directly related to the deflection potential $\Psi$ through
\begin{equation}
\kappa = \frac{1}{2}\Delta_{\vec\theta}\Psi
\label{eq:0d}
\end{equation}
and the shear components $\gamma_{1}$ and $\gamma_{2}$ are given by
\begin{equation}
\begin{split}
\gamma_{1} &= \frac{1}{2}\left(\frac{\partial^{2}\Psi}{\partial\theta_{1}^{2}}-\frac{\partial^{2}\Psi}{\partial\theta_{2}^{2}}\right),\quad\gamma_{2} = \frac{\partial^{2}\Psi}{\partial\theta_{1}\partial\theta_{2}},\\
\gamma &= \sqrt{\gamma_{1}^{2}+\gamma_{2}^{2}}.
\end{split}
\label{eq:0e}
\end{equation}
Because of Liouville's theorem, gravitational lensing preserves the surface brightness, but it changes the apparent solid angle of a source. The resulting flux ratio between image and source can be expressed in terms of the amplification $A$,
\begin{equation}
A^{-1} = (1-\kappa)^{2}-\gamma^{2}.
\label{eq:0f}
\end{equation}
Considering the symmetry properties of our cylindrical lens model and the configuration in Figure \ref{fig1}, equation \eqref{eq:0d} further simplifies to
\begin{equation}
\kappa (y) = \frac{1}{2}\frac{D_{l}D_{ls}}{D_{s}}\frac{{\partial\hat\alpha (y)}}{{\partial y}},
\label{eq:0g}
\end{equation}
with the convergence $\kappa$ being related to the quantities $\gamma$ ($\gamma^{2}=\gamma_{1}^{2}$ and $\gamma_{2}=0$) and $A$ as follows:
\begin{equation}
\kappa = \gamma =\frac{1-A^{-1}}{2}.
\label{eq:0h}
\end{equation}
Furthermore, let us introduce the complex reduced shear $g$ given by
\begin{equation}
g = \frac{\gamma_{1}+i\gamma_{2}}{1-\kappa}.
\label{eq:0i}
\end{equation}
This quantity is the expectation value of the ellipticity $\chi$ of galaxies weakly distorted by the lensing effect, thus corresponding to the signal which can actually be observed. In our case, we find that the absolute value of the reduced shear is $|g|=\gamma/(1-\kappa)$, and assuming that $\kappa=\gamma\ll 1$, we obtain $|g|\sim\kappa=\gamma$.
Note that the above result is independent of the particular law of gravity.
\subsection{Newtonian Case}
The Newtonian gravitational field of our filament model is given by
\begin{equation}
g_N (r) = \lvert\vec\nabla\Phi_{N}(r)\rvert = \left\{
  \begin{array}{ll}
    {\dfrac{{G\lambda}}{{2\pi}}\dfrac{r}{{R_f^2 }},} &  \hbox{$r < R_f$,} \\ &\\
    {\dfrac{{G\lambda}}{{2\pi}}\dfrac{1}{r},} &  \hbox{$r \ge R_f$,} \\
  \end{array}\right .
\label{eq:1}
\end{equation}
with $\lambda$ being the previously defined line density given by equation \eqref{eq:linedef}.
\begin{figure*}
 \centering
   \begin{minipage}[t]{8.5 cm}
\begin{center} 
\includegraphics[trim=43 0 0 0,width=0.85\textwidth]{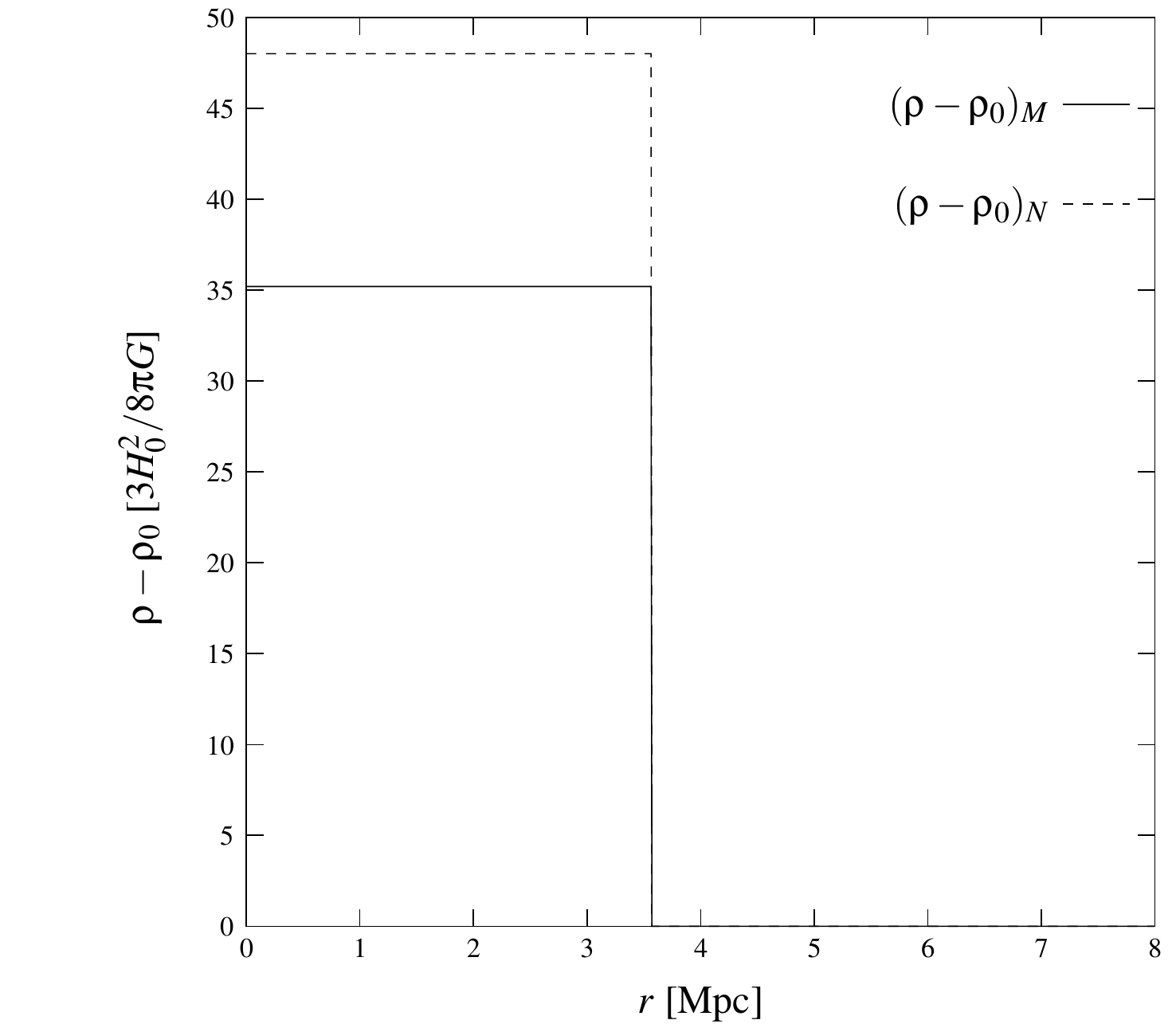}\\[0.3cm]
\includegraphics[trim=20 0 0 0,width=0.85\textwidth]{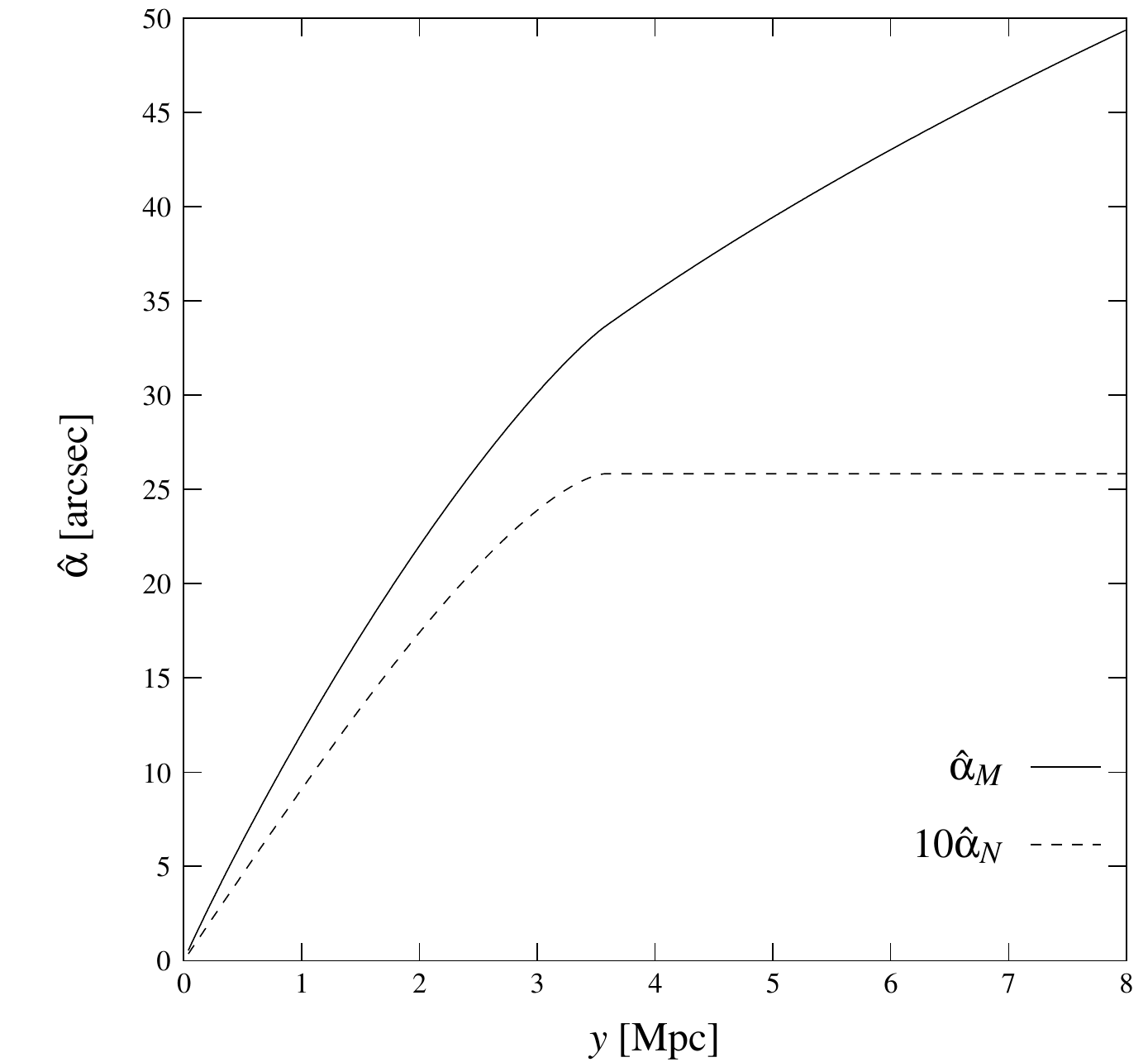}
\end{center}
  \end{minipage}
\qquad
 \begin{minipage}[t]{8.5 cm}
\begin{center}
\includegraphics[trim=20 0 0 0,width=0.85\textwidth]{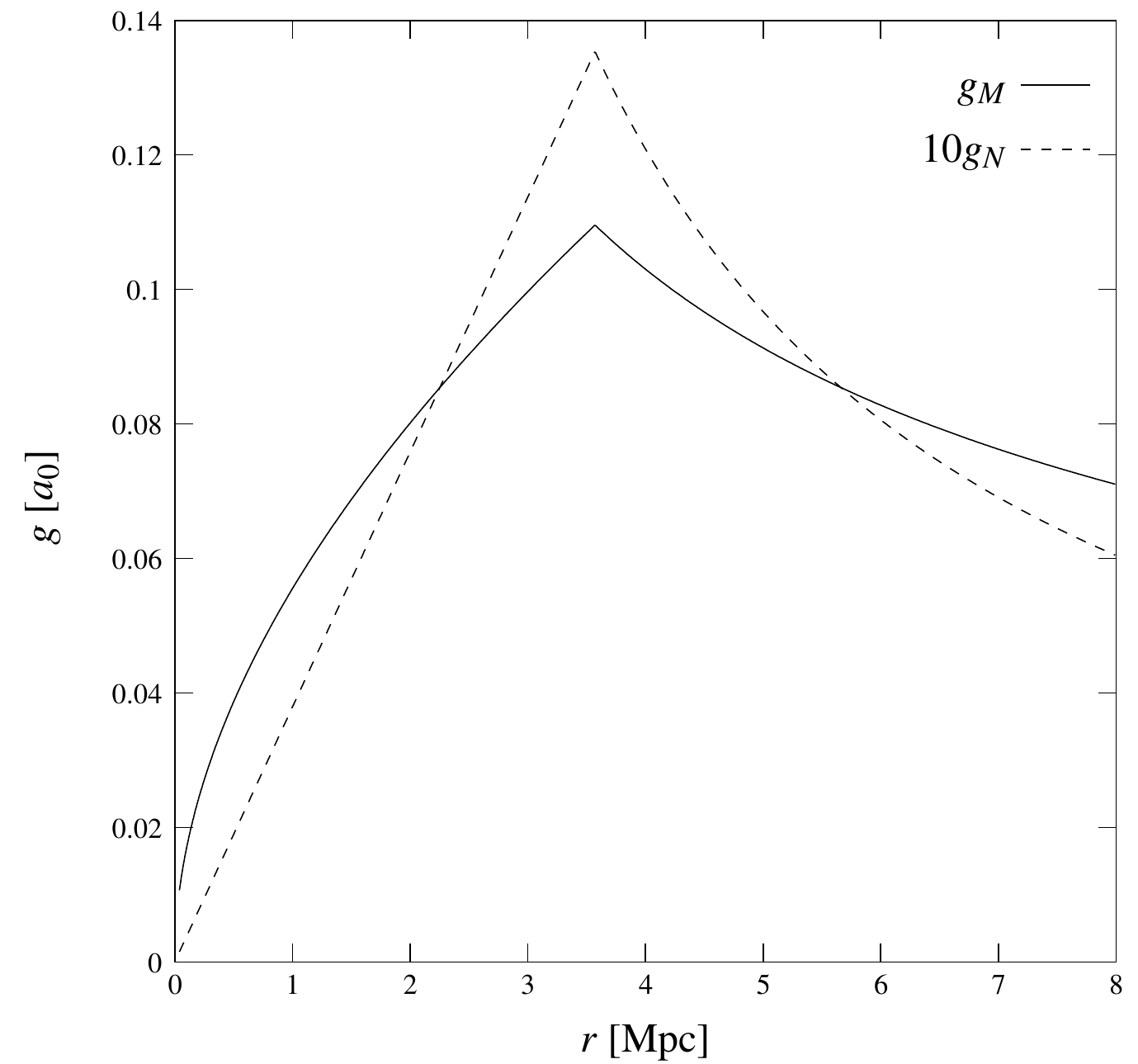}\\[0.3cm]
\includegraphics[trim=20 0 0 0,width=0.85\textwidth]{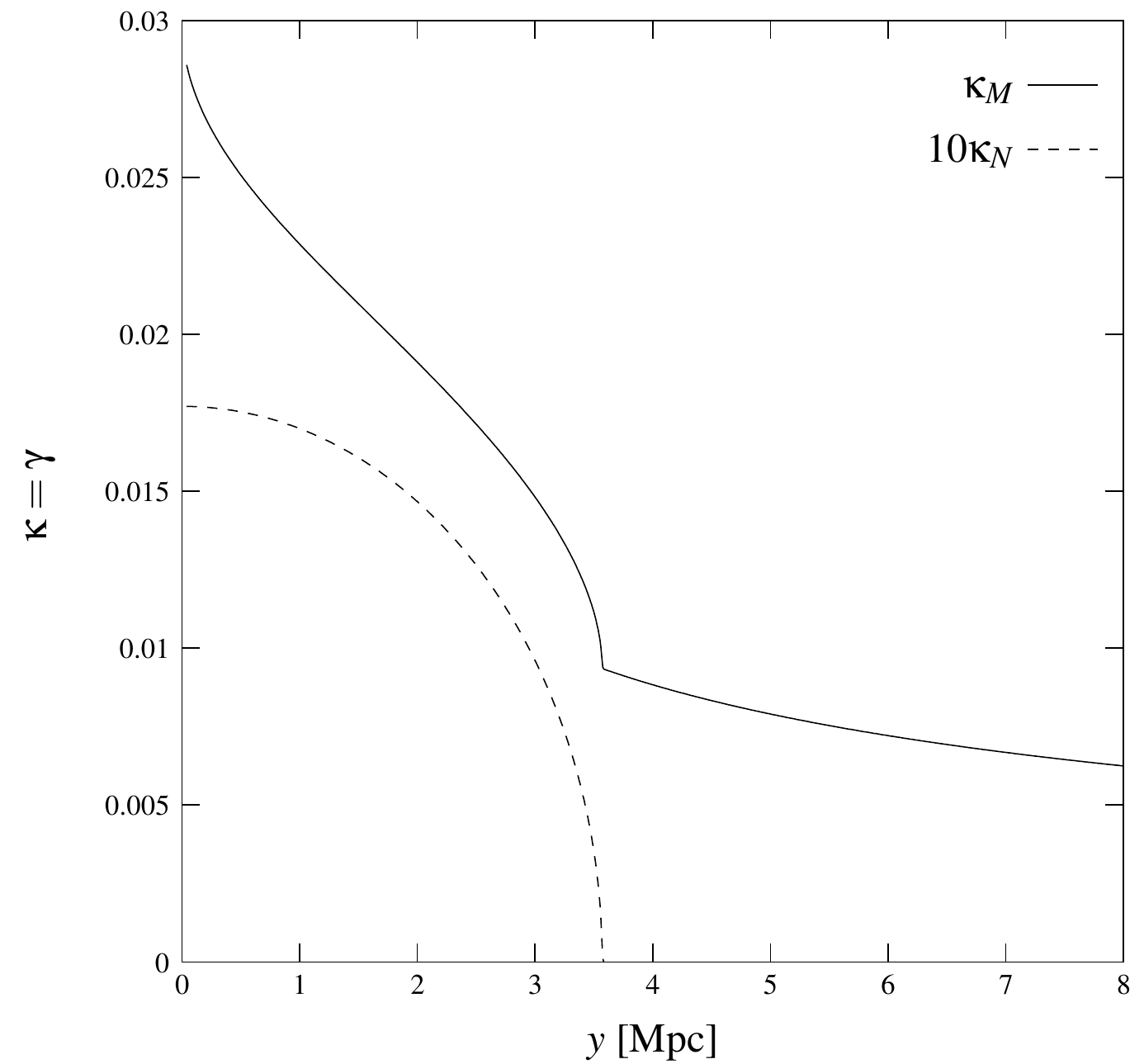}
\end{center}
\end{minipage}
\caption{Density profile $\rho(r)$ ({\it top left}), radial evolution $g(r)$ of the total gravitational acceleration ({\it top right}), deflection angle $\hat\alpha(y)$ ({\it bottom left}) and convergence $\kappa(y)$ ({\it bottom right}; $\kappa=\gamma=(1-A^{-1})/2$) in Newtonian ({\it dashed line}) and MONDian ({\it solid line}) gravity for the uniform filament cylinder model whose axis is inclined by an angle $\theta=90^{\circ}$ to the line of sight, assuming $z_{l}=1$, $z_{s}=3$, and the flat $\mu$HDM cosmology of eq. \eqref{eq:10} in MOND/TeVeS. The radius of the filament is $R_{f}=2.5$ $h^{-1}$ Mpc, and the overdensity within the filament is taken as $20$ times the mean density $\rho_{0}$ in accordance with the results of Colberg et al. (2005). Note that, for consistency, the Newtonian results are based on a flat $\Lambda$CDM cosmology with $\Omega_{m}=0.3$ and $\Omega_{\Lambda}=0.7$.}
\label{fig2a}
    \end{figure*}
\begin{figure*}
 \centering
   \begin{minipage}[t]{8.5 cm}
\begin{center} 
\includegraphics[trim=43 0 0 0,width=0.85\textwidth]{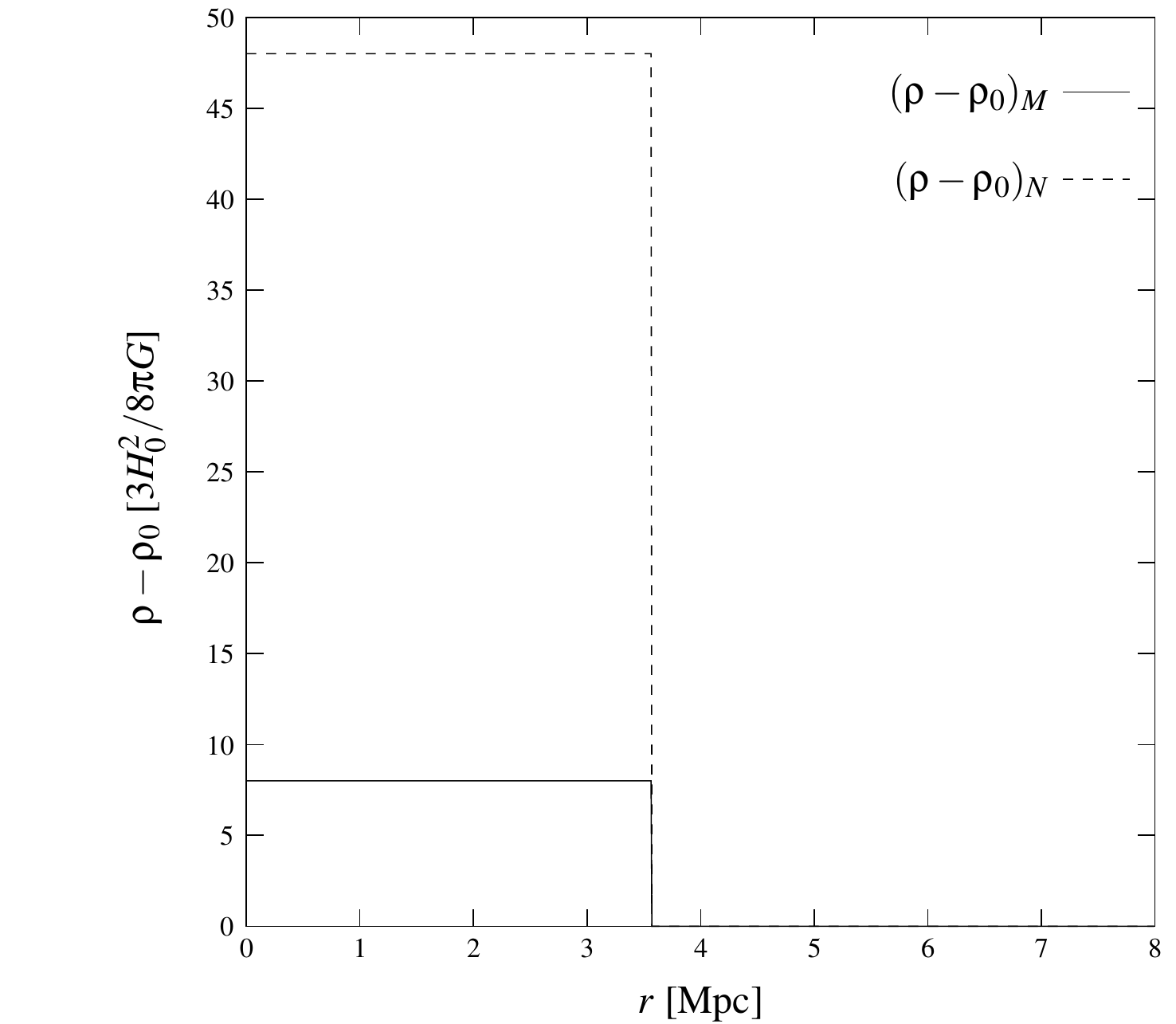}\\[0.3cm]
\includegraphics[trim=20 0 0 0,width=0.85\textwidth]{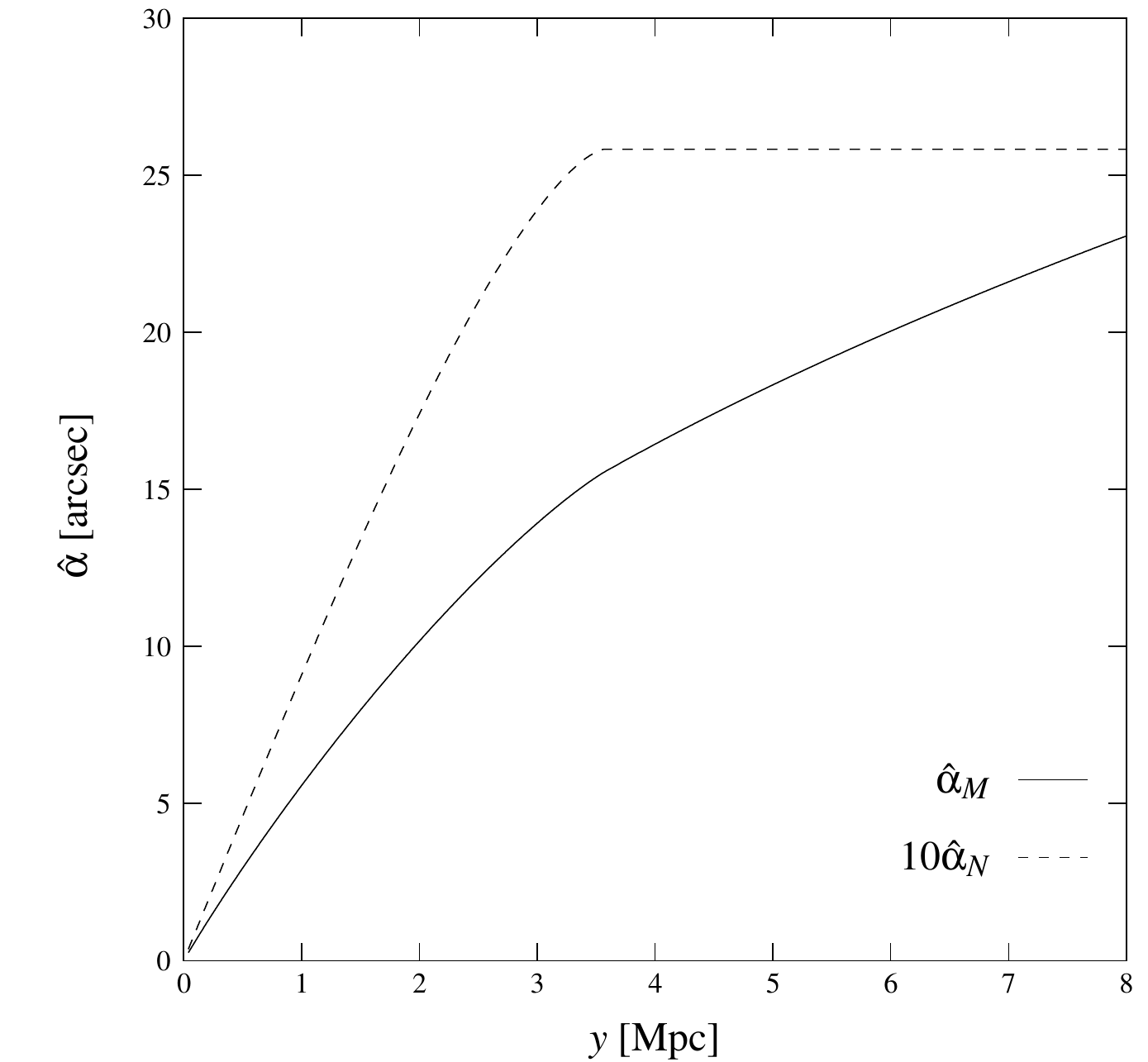}
\end{center}
  \end{minipage}
\qquad
 \begin{minipage}[t]{8.5 cm}
\begin{center}
\includegraphics[trim=20 0 0 0,width=0.85\textwidth]{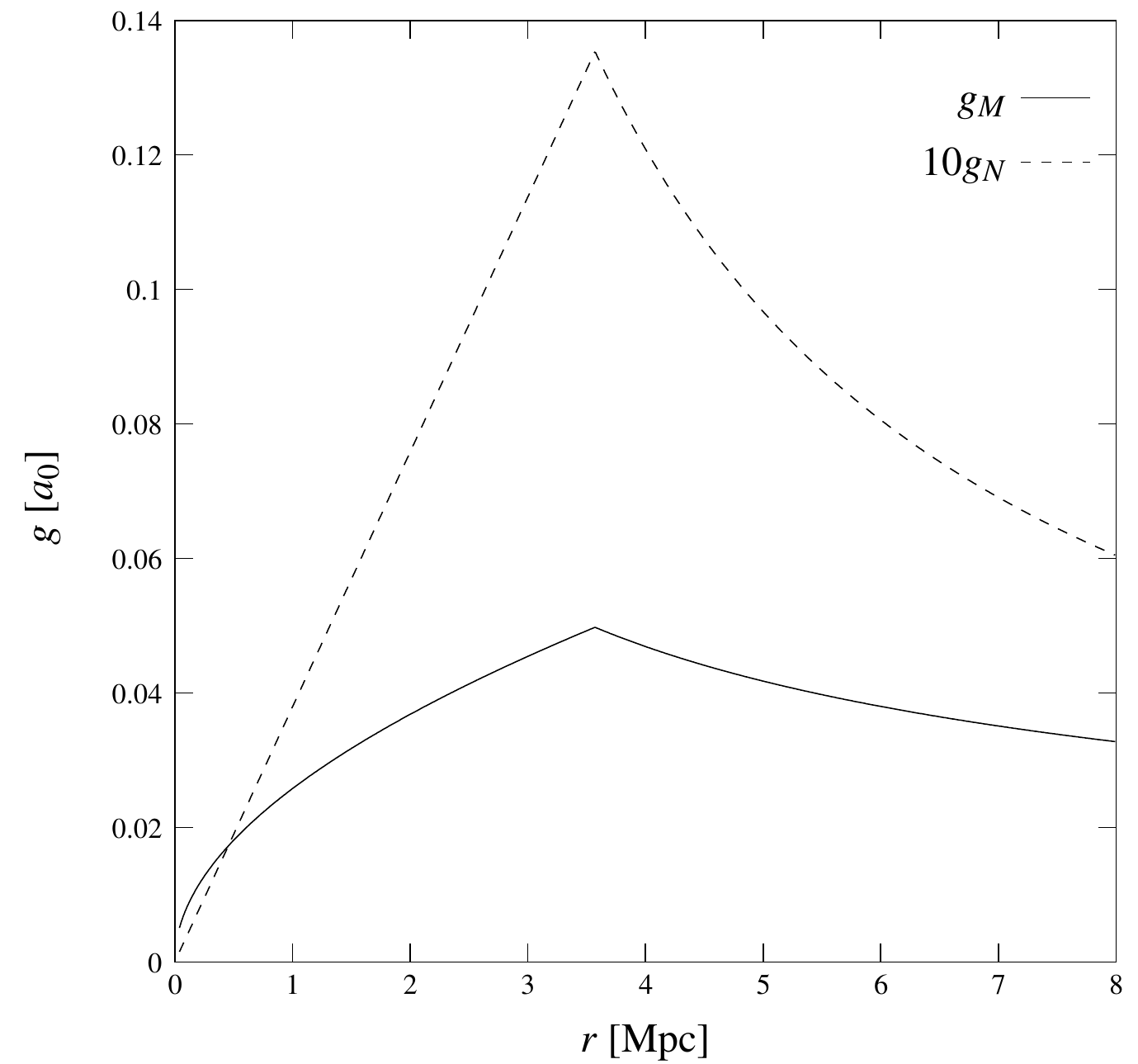}\\[0.3cm]
\includegraphics[trim=20 0 0 0,width=0.85\textwidth]{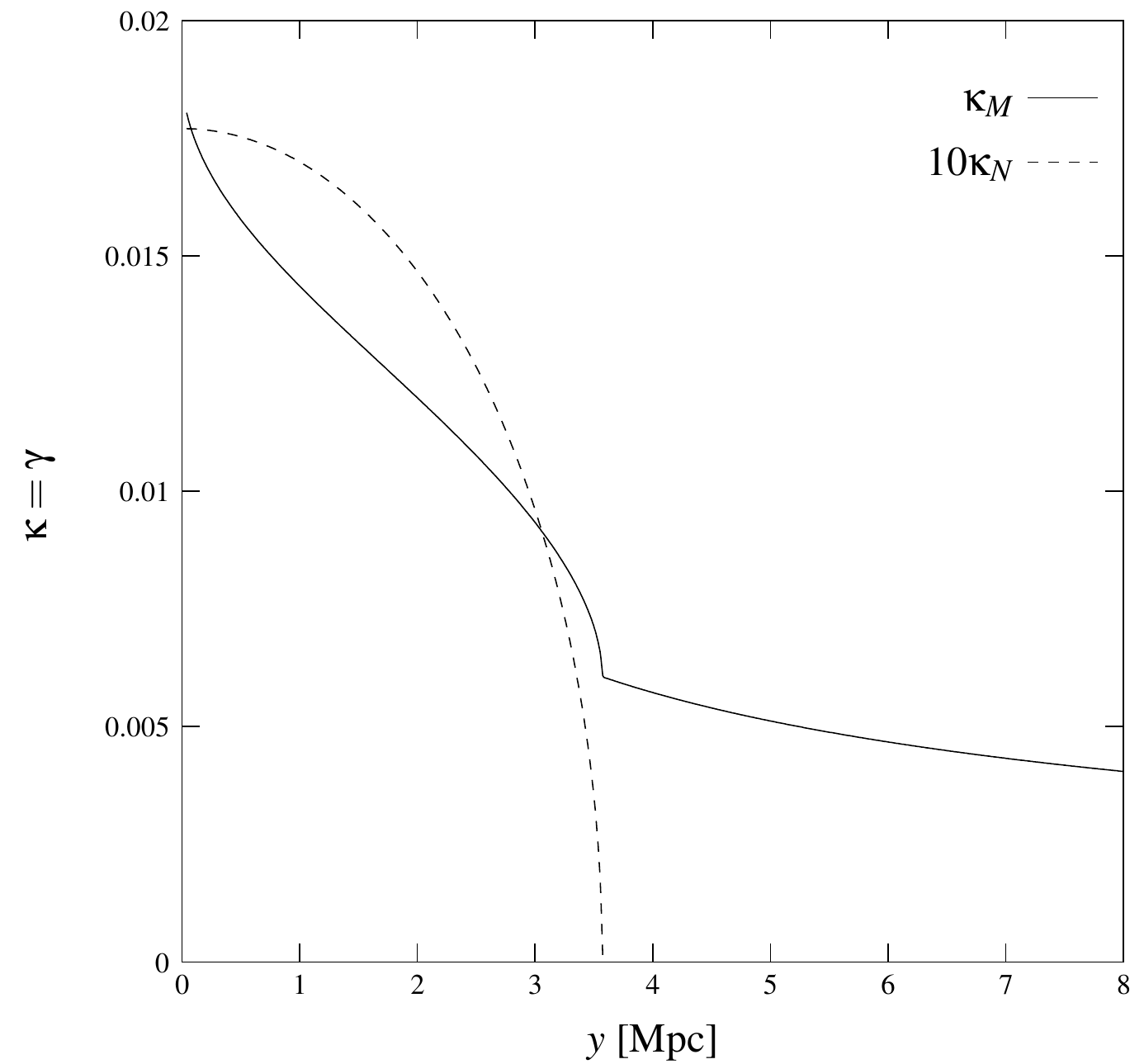}
\end{center}
\end{minipage}
\caption{Density profile $\rho(r)$ ({\it top left}), radial evolution $g(r)$ of the total gravitational acceleration ({\it top right}), deflection angle $\hat\alpha(y)$ ({\it bottom left}) and convergence $\kappa(y)$ ({\it bottom right}; $\kappa=\gamma=(1-A^{-1})/2$) in Newtonian ({\it dashed line}) and MONDian ({\it solid line}) gravity for the uniform filament cylinder model whose axis is inclined by an angle $\theta=90^{\circ}$ to the line of sight, assuming $z_{l}=1$, $z_{s}=3$, and the flat minimal-matter cosmology of eq. \eqref{eq:10a} in MOND/TeVeS. The radius of the filament is $R_{f}=2.5$ $h^{-1}$ Mpc, and the overdensity within the filament is taken as $20$ times the mean density $\rho_{0}$ in accordance with the results of Colberg et al. (2005). Note that, for consistency, the Newtonian results are based on a flat $\Lambda$CDM cosmology with $\Omega_{m}=0.3$ and $\Omega_{\Lambda}=0.7$.}
\label{fig2b}
    \end{figure*}

\textbf{(I)} For $ R_f \le y$, evaluating integral \eqref{eq:0a} yields
\begin{equation}
\hat\alpha_N (y) = G\lambda ={\rm const}.
\label{eq:1a}
\end{equation}
Inserting the above into equation \eqref{eq:0g}, we may obtain the corresponding convergence field. As expected,
$\kappa_{N}$ equals zero outside the cylinder's projected matter density.

\textbf{(II)} For $ y< R_f$, the deflection angle has to be calculated from
\begin{equation}
\hat\alpha_N (y) = \frac{2G\lambda y}{\pi}\left\{\int\limits_{y}^{R_{f}} { {\frac{rdr}{R_{f}^{2}\sqrt{r^{2}-y^{2}}}}
}   + \int\limits_{R_{f}}^{\infty} {
{\frac{dr}{r \sqrt{r^{2}-y^{2}}}} }  \right\}.
\label{eq:2}
\end{equation}
Carrying out the integrations in equation \eqref{eq:2}, we finally end up with the following expression:
\begin{equation}
\hat\alpha_N (y) = \frac{2G\lambda}{\pi}\left (\dfrac{y\sqrt{R_{f}^{2}-y^{2}}}{R_{f}^{2}}+\arcsin\left (\dfrac{y}{R_{f}}\right )\right ).
\label{eq:3}
\end{equation}
Using equation \eqref{eq:0g}, the convergence in this case turns out to be
\begin{equation}
\kappa_N (y) = 2\frac{{D_l D_{ls} }}{{D_s }}\frac{G\lambda}{{\pi R_f^2 }}\sqrt
{R_f^2  - y^2 }.
\label{eq:4}
\end{equation}

\subsection{MONDian Case}
Now we shall consider light deflection within the framework of TeVeS gravity. Choosing a certain smooth form of
the free interpolating function $\mu$ \citep[for further details see][]{teves} which has been used by \cite{lenstest},
the total gravitational acceleration may be written in the following way:
\begin{equation}
g_M (r) = \lvert\vec\nabla\Phi_{M}(r)\rvert = g_N (r) + \sqrt {g_N (r)a_0 },
\label{eq:5}
\end{equation}
with $r$ again being the cylindrical radial coordinate and $\Phi_{M}(r)$ the total non-relativistic gravitational potential in TeVeS. The constant $a_0  = 1.2\times 10^{-10}$ m s$^{-2}$ characterizes the acceleration scale at which MONDian effects start to become important compared to Newtonian contributions. Since filaments are the most low-density structures within the universe, their internal (Newtonian) gravity is very small. Therefore, the MONDian influence yields an enhancement of the gravitational field which is on the order of $a_{0}/g_{N}$, being extremely large in such objects. For this reason, we may expect a substantial difference concerning the lensing signal caused by filamentary structures in TeVeS.
Equipped with equations \eqref{eq:0a}, \eqref{eq:1} and \eqref{eq:5} we are ready to proceed with the analysis of our cylindrical filament model:

\textbf{(I)} For $ R_f \le y$, the deflection angle is given by

\begin{equation}
\begin{split}
\hat\alpha _M (y) &= \hat\alpha _N (y) +\sqrt{\dfrac{8G\lambda a_{0}}{\pi}}y\int\limits_{y}^{\infty}\frac{dr}{\sqrt{r}\sqrt{r^{2}-y^{2}}}
\\
&= G\lambda + \frac{{\Gamma (1/4)}}{{\Gamma (3/4)}}\sqrt{2G\lambda a_{0}y}.
\end{split}
\label{eq:6}
\end{equation}
In this case, the convergence reads as follows:
\begin{equation}
\kappa _M (y) = \frac{{D_l D_{ls} }}{{D_s }}\frac{{\Gamma (1/4)}}{{\Gamma (3/4)}}\sqrt{\frac{{G\lambda a_{0}}}{{8y}}}.
\label{eq:7}
\end{equation}

\textbf{(II)} For $ y< R_f$, integral \eqref{eq:0a} has to be split in several parts, similarly to equation \eqref{eq:2}. Using elementary calculus, we finally arrive at
\begin{equation}
\begin{split}
&\hat\alpha _M (y) = \hat\alpha_{N}(y)\\
& + \sqrt{\dfrac{2G\lambda a_{0}}{\pi}}\dfrac{y^{3/2}}{R_{f}}\left\lbrack 4\sqrt{\dfrac{R_{f}^{2}-y^{2}}{R_{f}y}}-\mathcal{B}_{\left (y^{2}/R_{f}^{2},1\right )}\left (\frac{3}{4},\frac{1}{2}\right )\right\rbrack \\
& +\sqrt{\dfrac{2G\lambda a_{0}y}{\pi}}\mathcal{B}_{\left (0,y^{2}/R_{f}^{2}\right )}\left (\frac{1}{4},\frac{1}{2}\right ),
\end{split}
\label{eq:8}
\end{equation}
where $\hat\alpha_{N}(y)$ is the Newtonian deflection angle given by equation \eqref{eq:3} and $\mathcal{B}_{(p,q)}(a,b)$ is the generalized incomplete Beta function defined by
\begin{equation}
\mathcal{B}_{(p,q)}(a,b) = \int\limits_{p}^{q}t^{a-1}(1-t)^{b-1}dt, \quad Re(a),Re(b)>0.
\label{eq:9}
\end{equation}
As the expression for the convergence $\kappa_{M}$ turns out to be quite lengthy, we will drop it at this point.

From equations \eqref{eq:6} and \eqref{eq:7}, we find that the deflection angle $\alpha_{M}$ outside the cylinder's projection increases with the square root of the impact parameter $y$ ($\alpha_{N}=const$), the convergence $\kappa_{M}$ decreases with the inverse square root of $y$ ($\kappa_{N}=0$). In fact, this reveals quite a fundamental difference between MOND/TeVeS and GR; since $\kappa_{N}=0$, we also have $\gamma_{N}=0$ and $A_{N}=1$ according to equation \eqref{eq:0h},   meaning that there will be no distortion effects, as well as no change in the total flux between source and image, i.e. wherever the projected matter density is zero, the lens mapping will turn into identity.
However, this is no longer true in the context of MOND/TeVeS as the convergence and the shear field do not vanish (see Fig. \ref{fig2a}). Obviously, this is a case where the MONDian influence does not only enhance effects that are already present in GR, but rather creates something new, which, in principle, could be used to distinguish between laws of gravity (see \S \ref{discussion}).

Varying the inclination angle $\theta$ of the filament's axis to the line of sight, the lensing properties derived in this section have to be rescaled by a factor of $\sin^{-1}\theta$ in both GR and TeVeS.

\section{Model Application}
\label{app}
In their $\Lambda$CDM large-scale structure simulation, \cite{LCDMfilament} have shown that there are
close cluster pairs with a separation of $5 h^{-1}$ Mpc or less which are always
connected by a filament. At separations between $15$ and $20$ $h^{-1}$ Mpc, still about a third
of cluster pairs are connected by a filament. On average, more massive clusters are connected
to more filaments than less massive ones. In addition, the current simulation indicates that the
most massive clusters form at the intersections of the filamentary backbone of large-scale
structure. For straight filaments, the radial profiles show a fairly well-defined radius $R_f$
beyond which the profiles closely follow an $r^{-2}$ power law, with $R_f$ being around $2.0$ $h^{-1}$ Mpc
for the majority of filaments. The enclosed overdensity within $R_f$ varies from
a few times up to $25$ times the mean density, independent of the filament's length. Along
the filaments' axes, material is not distributed uniformly. Towards the clusters, the density
rises, indicating the presence of the cluster infall regions.
\begin{figure*}
 \centering
   \begin{minipage}[t]{8.5 cm}
\begin{center} 
\includegraphics[trim=43 0 0 0,width=0.85\textwidth]{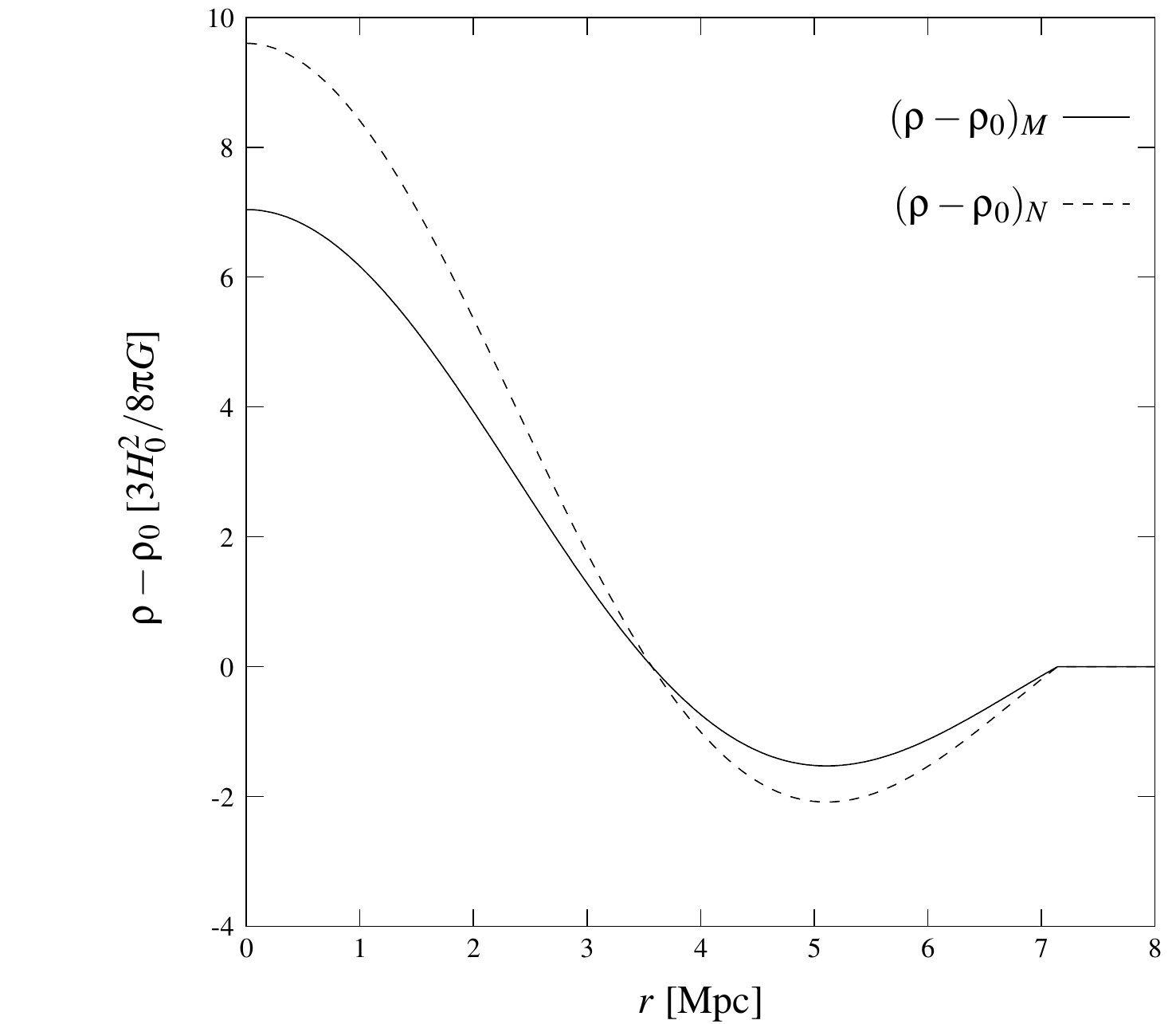}\\[0.3cm]
\includegraphics[trim=20 0 0 0,width=0.85\textwidth]{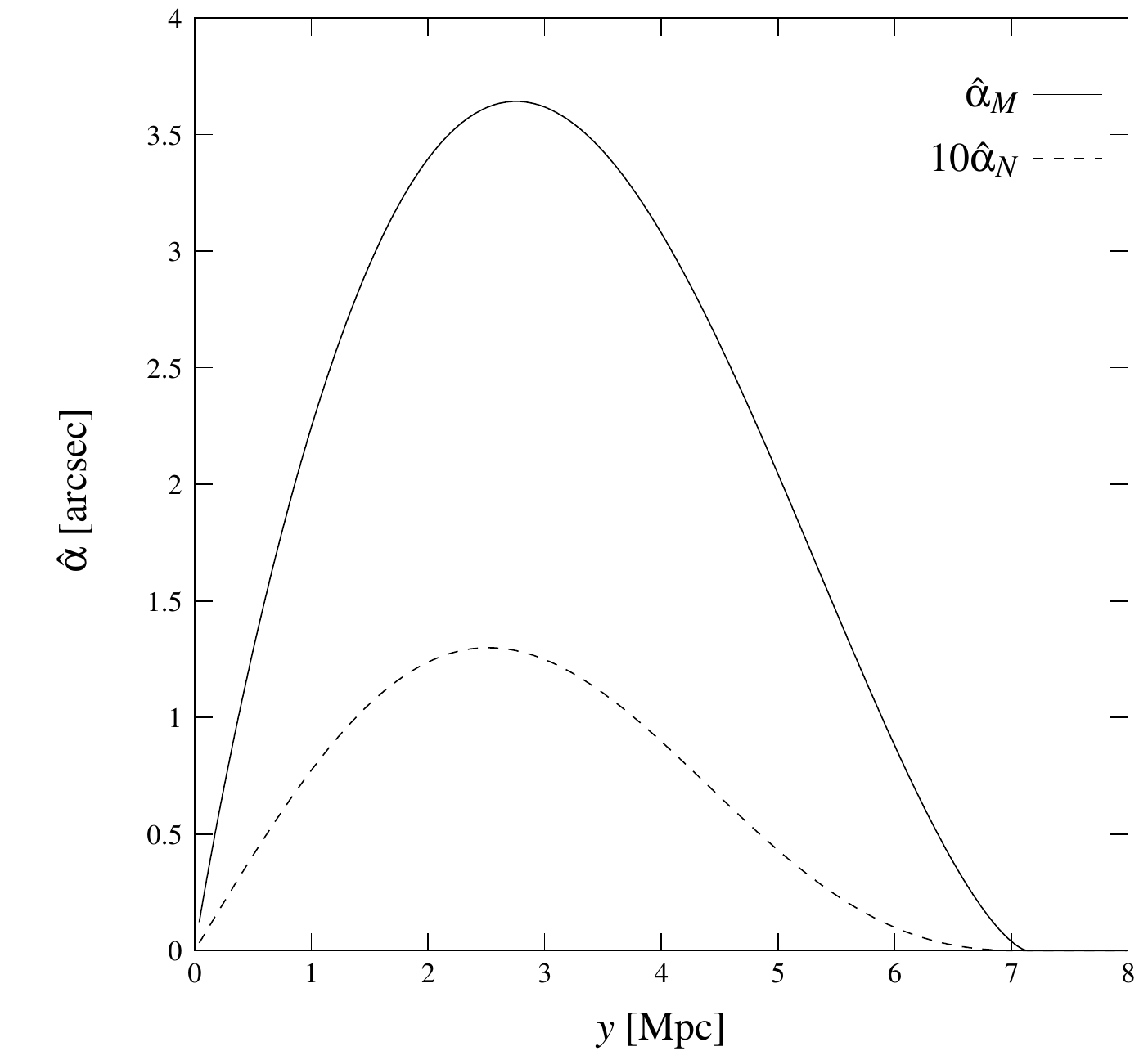}
\end{center}
  \end{minipage}
\qquad
 \begin{minipage}[t]{8.5 cm}
\begin{center}
\includegraphics[trim=20 0 0 0,width=0.85\textwidth]{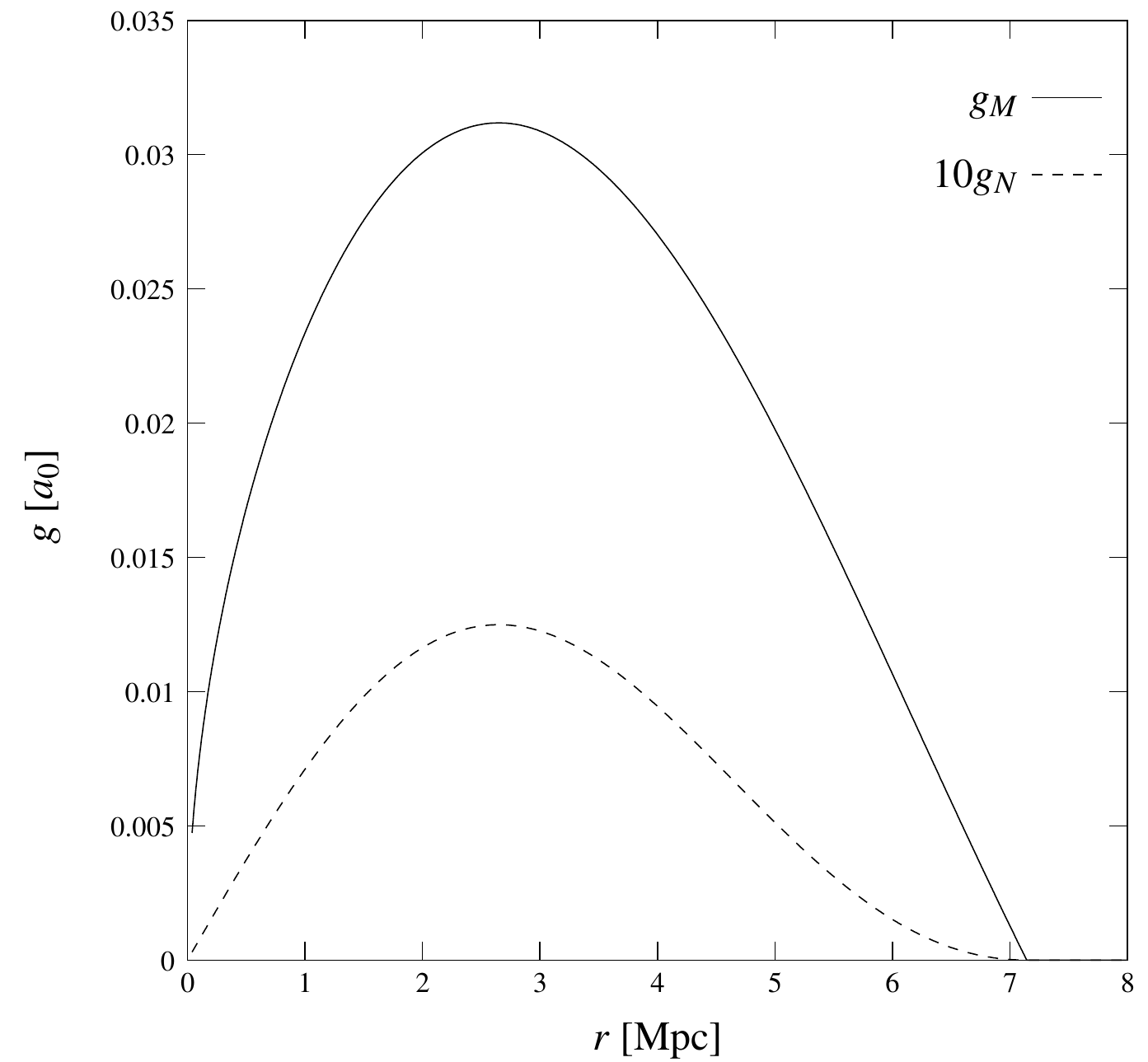}\\[0.3cm]
\includegraphics[trim=20 0 0 0,width=0.85\textwidth]{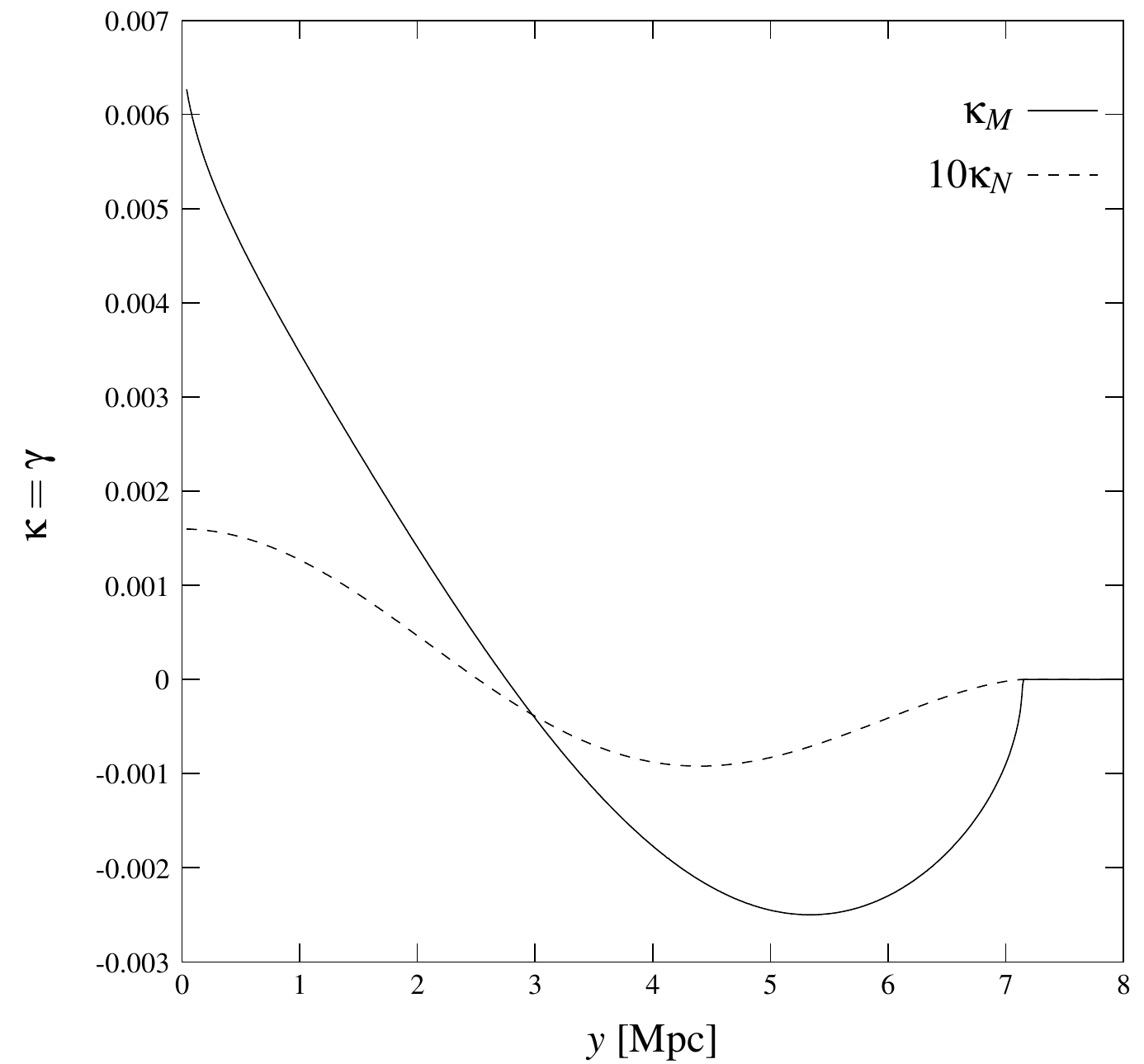}
\end{center}
\end{minipage}
\caption{Density profile $\rho(r)$ ({\it top left}), radial evolution $g(r)$ of the total gravitational acceleration ({\it top right}), deflection angle $\hat\alpha(y)$ ({\it bottom left}) and convergence $\kappa(y)$ ({\it bottom right}; $\kappa=\gamma=(1-A^{-1})/2$) in Newtonian ({\it dashed line}) and MONDian ({\it solid line}) gravity for the oscillating density model given by eq. \eqref{eq:12} ($\theta=90^{\circ}$), assuming $\delta_{0}=4$, $R_{f}=2.5$ $h^{-1}$ Mpc, $z_{l}=1$, $z_{s}=3$, and the flat $\mu$HDM cosmology in eq. \eqref{eq:10} for MOND/TeVeS. Note that, for consistency, the Newtonian results are based on a flat $\Lambda$CDM cosmology with $\Omega_{m}=0.3$ and $\Omega_{\Lambda}=0.7$.}
\label{fig3}
\end{figure*}
\begin{figure*}
 \centering
   \begin{minipage}[t]{8.5 cm}
\begin{center} 
\includegraphics[trim=43 0 0 0,width=0.85\textwidth]{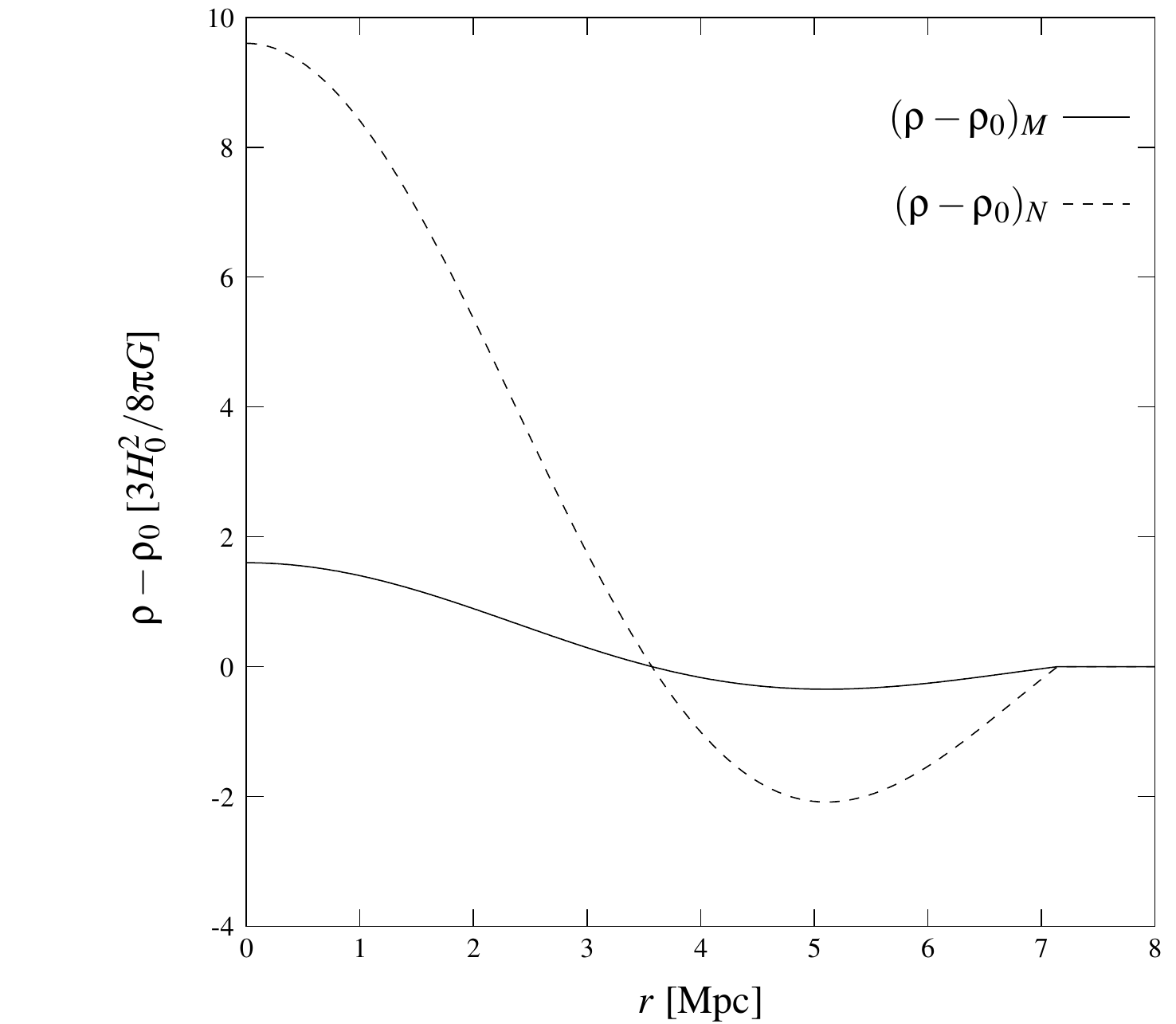}\\[0.3cm]
\includegraphics[trim=20 0 0 0,width=0.85\textwidth]{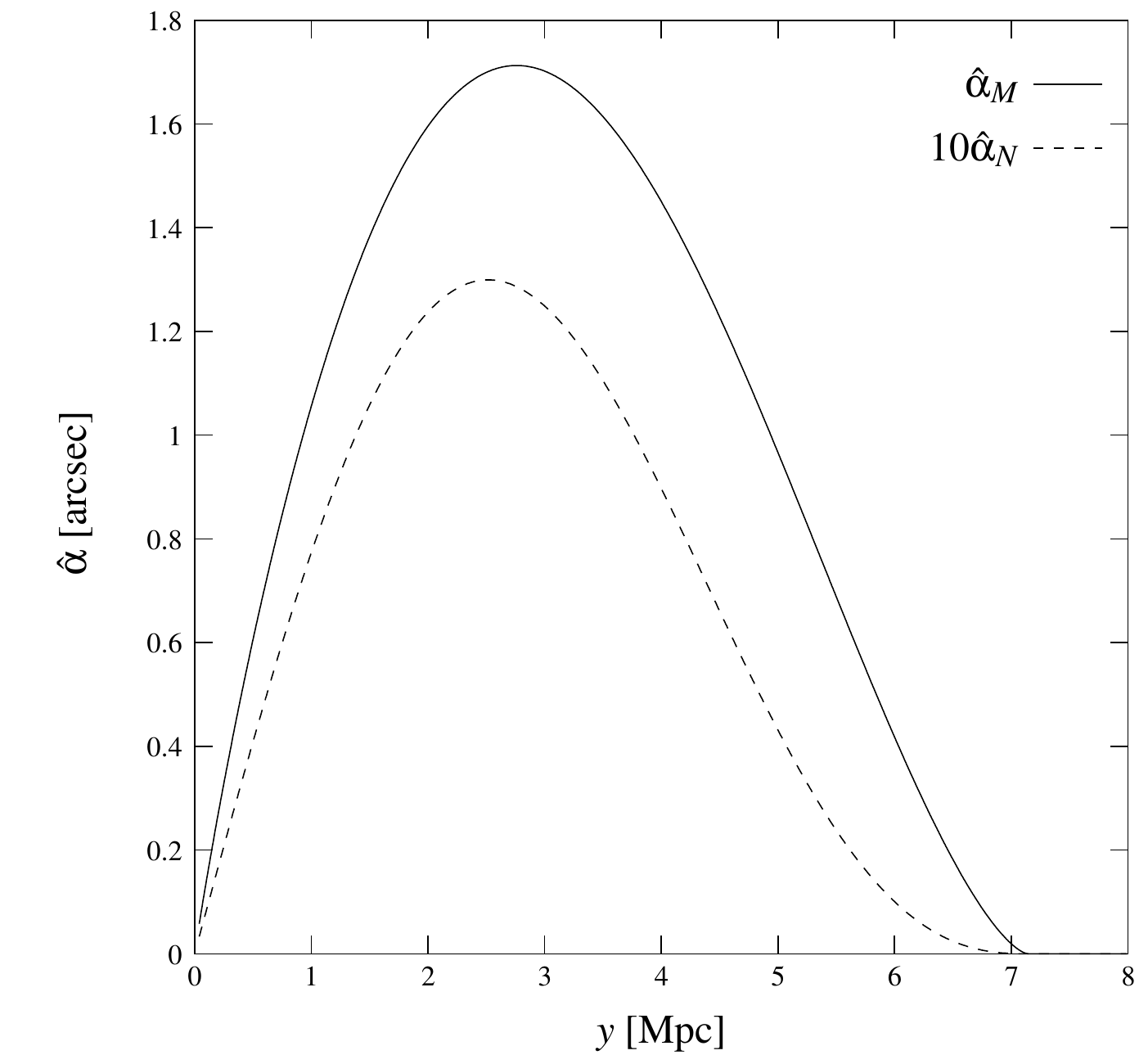}
\end{center}
  \end{minipage}
\qquad
 \begin{minipage}[t]{8.5 cm}
\begin{center}
\includegraphics[trim=20 0 0 0,width=0.85\textwidth]{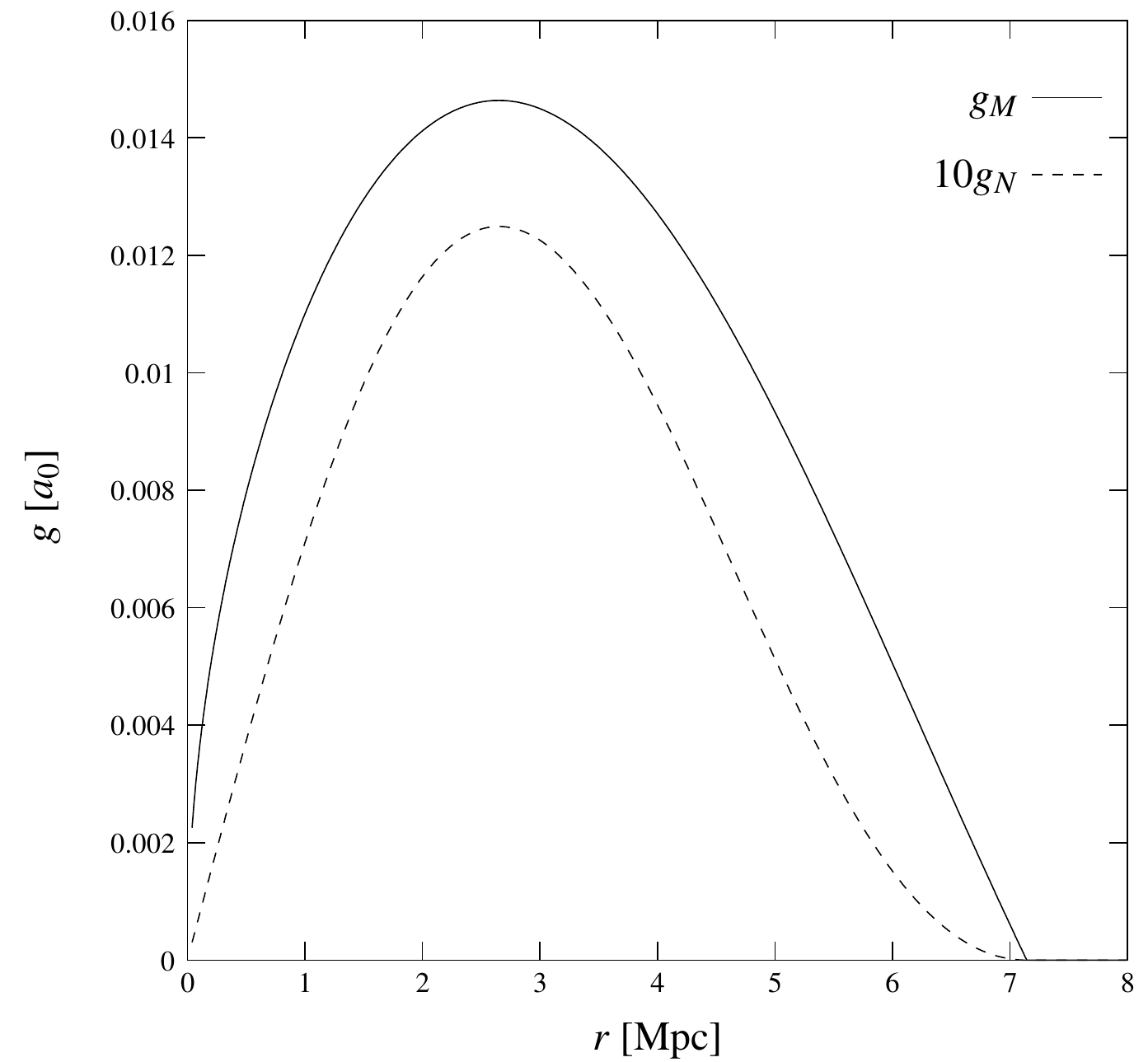}\\[0.3cm]
\includegraphics[trim=20 0 0 0,width=0.85\textwidth]{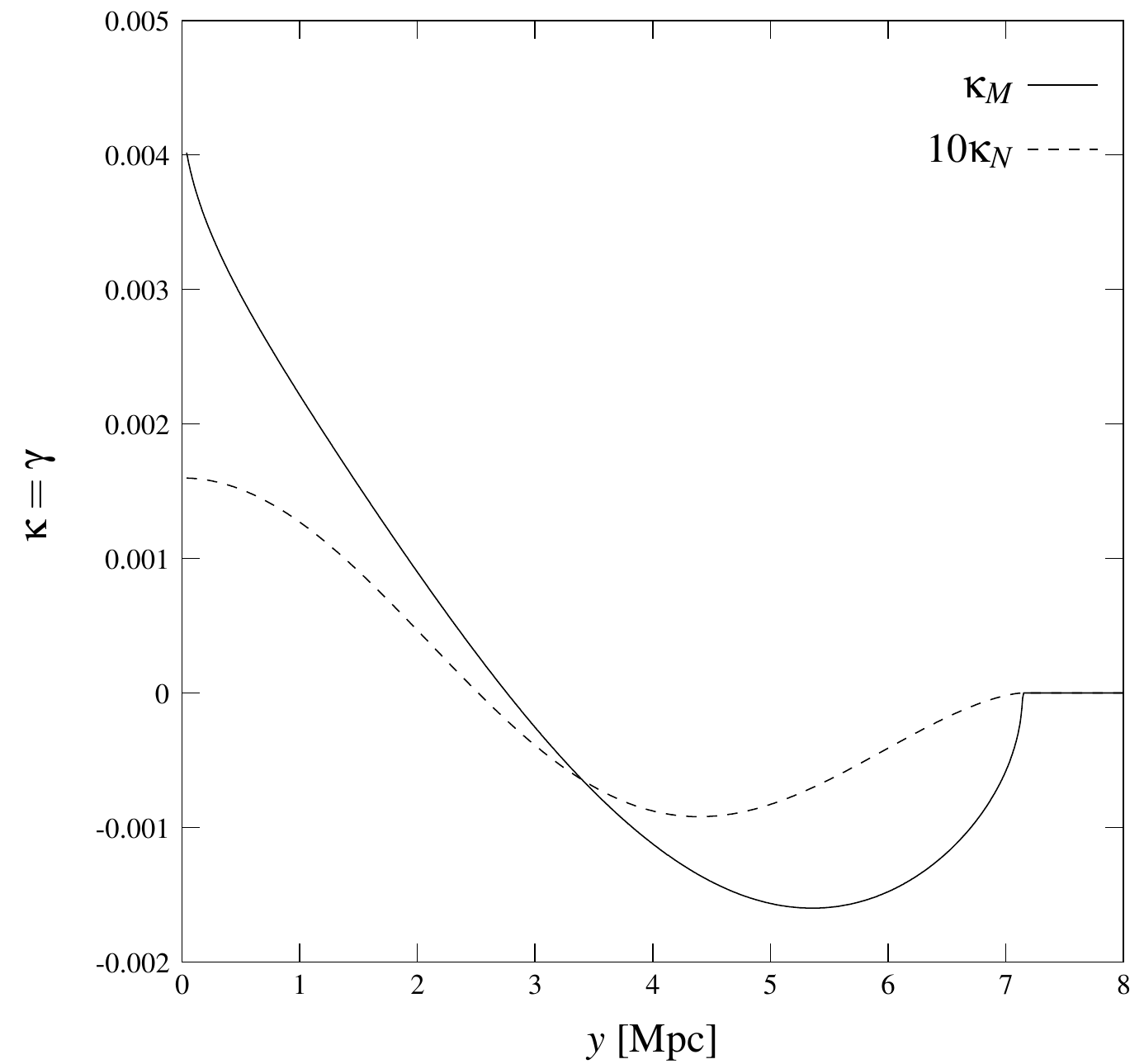}
\end{center}
\end{minipage}
\caption{Density profile $\rho(r)$ ({\it top left}), radial evolution $g(r)$ of the total gravitational acceleration ({\it top right}), deflection angle $\hat\alpha(y)$ ({\it bottom left}) and convergence $\kappa(y)$ ({\it bottom right}; $\kappa=\gamma=(1-A^{-1})/2$) in Newtonian ({\it dashed line}) and MONDian ({\it solid line}) gravity for the oscillating density model given by eq. \eqref{eq:12} ($\theta=90^{\circ}$), assuming $\delta_{0}=4$, $R_{f}=2.5$ $h^{-1}$ Mpc, $z_{l}=1$, $z_{s}=3$, and the flat minimal-matter cosmology in eq. \eqref{eq:10a} for MOND/TeVeS. Note that, for consistency, the Newtonian results are based on a flat $\Lambda$CDM cosmology with $\Omega_{m}=0.3$ and $\Omega_{\Lambda}=0.7$.}
\label{fig4}
    \end{figure*}
    
As previously stated, we will assume that filamentary structures have similar properties in MOND/TeVeS and in a CDM dominated universe based on GR. Our assumption is based on the $\mu$HDM cosmology (see \S \ref{intro}) and on the fact that filaments are generic\footnotemark\footnotetext[7]{Note that the occurrence of filamentary structures is a generic feature of gravitational collapse from a Gaussian random field which does not depend on the specific form of the law of gravity.} and have similar characteristics in hot dark matter (HDM) and CDM scenarios \citep{knebe1,knebe2,wdm}. For instance, neutrinos are known to collapse into sheets and filaments in HDM simulations. Concerning our filament model introduced in
Sec. \ref{model}, we therefore take the filament's radius as $R_{f}=2.5$ $h^{-1}$ Mpc, and taking the overdensity within the filament to be $20$ times the intergalactic mean density $\rho_{0}$, we set $\delta=20$, with $\delta$ being the density contrast defined by
\begin{equation}
\delta = \frac{\rho-\rho_{0}}{\rho_{0}}.
\label{eq:contrast}
\end{equation}
Again, note that choosing the $\mu$HDM cosmology implies that filaments do not solely consist of baryonic matter but need an additional matter component, i.e. neutrinos, within the MOND paradigm, which is inferred from the previously mentioned discrepancies between dynamical and visible mass on galaxy cluster scale \citep{neutrinos2} as well as from the need for such a component to explain the CMB \citep{tevesneutrinocosmo}.

On the other hand, analyzing the Perseus-Pisces segment, \cite{MONDfilaments} concluded that a MONDian description of filaments would not need any additional non-baryonic mass component. Due to rather large systematic uncertainties, however, this result remains highly speculative and does not rule out our approach where filamentary structures have higher densities. Nevertheless, we will also include this case, where filaments consist of baryonic matter only, into our analysis. Since the absolute density of a filament in this situation is approximately by a factor $10-100$ smaller than in $\mu$HDM, we do expect the MONDian influence to become even more important. Encouraged by the MOND simulation of \cite{knebe3}, we shall stick to the assumption that both shapes and relative densities of filaments are similar to the CDM case when considering a universe made out of baryonic matter only, thus keeping the choice $\delta=20$.

In order to calculate the intergalactic mean density and the necessary angular diameter distances for lensing, we still need to set up a cosmological model in TeVeS. Depending on whether or not we assume massive neutrinos to be present in our universe, there will be a different cosmological background. If neutrinos are taken into account, then, for simplicity, we shall use the flat $\mu$HDM cosmology based on the parameters of \cite{tevesneutrinocosmo},
\begin{equation}
\Omega_{m}=0.22,\quad \Omega_{\Lambda}=0.78.
\label{eq:10}
\end{equation}
Considering a universe with baryons only, we choose a flat minimal-matter cosmology which is described by
\begin{equation}
\Omega_{m}=0.05,\quad \Omega_{\Lambda}=0.95.
\label{eq:10a}
\end{equation}
Furthermore, we shall set $h=0.7$ and calculate the model-dependent intergalactic mean density $\rho_{0}$ according to
\begin{equation}
\rho_{0} = \Omega_{m}\rho_{c}(1+z_{l})^{3},
\label{eq:10b}
\end{equation}
where $\rho_{c}=3H_{0}^{2}/8\pi G$ is the critical density and $z_{l}$ is the lens redshift, i.e. the filament's redshift. Although equation \eqref{eq:10a} has problems in explaining CMB observations due to its prediction of the last scattering sound horizon, both of the above cosmological models will be sufficient for assigning the distances of lenses and sources at redshifts $z \lesssim 3$ in the context of gravitational lensing. However, note that the simplicity of these models does not affect the upcoming analysis as we will limit ourselves to order-of-magnitude estimates only.

Concerning the framework of standard Newtonian gravity, we shall use a flat $\Lambda$CDM cosmology with $\Omega_{m}=0.3$ and $\Omega_{\Lambda}=0.7$, allowing a consistent comparison to our results in MOND/TeVeS.

\subsection{The $\mu$HDM Scenario}
\label{appne}
Using the cosmological parameters specified in equation \eqref{eq:10} within MOND/TeVeS and considering a filament which is inclined by an angle $\theta=90^{\circ}$ to the line of sight, both the Newtonian and the MONDian deflection angle as well as the corresponding convergence are plotted in the bottom left and bottom right panel of
Figure \ref{fig2a}, with the filament placed at redshift $z_{l}=1$ and background sources at $z_{s}=3$. Whereas the Newtonian signal is rather small, $\kappa_{N}\lesssim 10^{-3}$, the filament can create a convergence on the order of $\kappa\sim 0.01$ in MOND/TeVeS, even in the outer regions where $\kappa_{N}=0$ if we take into account that it can have other orientations, i.e. a different inclination angle $\theta$. For example, a nearly end-on filament, $\theta=10^\circ$, has a lensing power $6$ times larger than that of a face-on filament, $\theta=90^\circ$.

Using equation \eqref{eq:0h}, we therefore infer that a single MOND/TeVeS filament may generate a shear signal which is on the same order as the convergence, $\gamma\sim 0.01$, as well as an amplification bias at a $2\%$ level, $A^{-1}\sim 1.02$. In addition, we present the density $\rho(r)$ and the radial evolution of the total gravitational acceleration $g(r)$ for MOND and Newtonian dynamics in the top left and top right panel of Figure \ref{fig2a}, respectively. Note again that, for consistency, the Newtonian results are based on a flat $\Lambda$CDM cosmology with $\Omega_{m}=0.3$ and $\Omega_{\Lambda}=0.7$.
\subsection{The Baryons-only Scenario}
\label{appba}
Now let us switch to the baryonic cosmological background given by equation \eqref{eq:10a}. Keeping all remaining parameters exactly the same as in the last section, the corresponding results are presented in Figure \ref{fig2b}. Although the convergence is slightly smaller than in the $\mu$HDM case, roughly by a factor of $1.5-2$, we find that also in this case single filamentary structures are capable of producing a lensing signal which is of the same order, $\kappa\sim\gamma\sim 0.01$. Again, this is even true outside the ``edges" of the filament's projected matter density, accounting for the fact that the inclination angle $\theta$ may vary, $0^{\circ}\leq\theta\leq 90^{\circ}$.
\section{Oscillating Density Model}
\label{oscillator}
Matter density fluctuations are steadily present throughout the universe. Thus, as a more realistic approach, we shall use a fluctuating density profile to describe a filament and its surrounding area including voids, i.e. regions in the universe where the local matter density is below the intergalactic mean density.
To keep our analysis on a simple level, let us write the density fluctuation as ($r$ still denotes radial coordinate in cylindrical coordinates)
\begin{equation}
\delta(r) = \left\{
\begin{array}{ll}
    {\delta_0  \left (\dfrac{\pi r}{R_f}\right )^{-1}\sin\left (\dfrac{\pi r}{R_f}\right ),} &  \hbox{$r < 2R_f$,} \\ &\\
    {0,} &  \hbox{$r \ge 2R_f$,} \\
  \end{array}\right .
\label{eq:12}
\end{equation}
where $\delta(r)$ is the density contrast defined in equation \eqref{eq:contrast}, $\delta_0=4$ is the density fluctuation amplitude (this value ensures a positive overall matter density), and $R_{f}=2.5$ $h^{-1}$ Mpc is again the filament's characteristic radius. Multiplying with the mean density $\rho_{0}$ and integrating along the radial direction, we find that the mass per unit length enclosed by an infinite cylinder of radius $r$ reads as (note that we neglect the contribution due to the mean density background)
\begin{equation}
\frac{M(r)}{L} = \left\{
\begin{array}{ll}
    {\dfrac{2\rho_0 \delta_0 R_f^2}{\pi}\left\lbrack 1-\cos\left (\dfrac{\pi r}{R_f}\right )\right\rbrack ,} &  \hbox{$r < 2R_f$,} \\ &\\
    {0,} &  \hbox{$r \ge 2R_f$,} \\
  \end{array}\right .
\label{eq:13}
\end{equation}
where $\rho_0$ is the mean intergalactic matter density given by equation \eqref{eq:10b}.
From equation \eqref{eq:13}, we directly see that the Newtonian gravitational acceleration in this case is
\begin{equation}
g_N(r)=\frac{GM(r)}{2 \pi L}\frac{1}{r}.
\label{eq:14}
\end{equation}
Using equations \eqref{eq:0a}, \eqref{eq:5} and \eqref{eq:14}, we are now able to numerically calculate the lensing properties of this configuration. Choosing lens and source redshift again as $z_l=1$ and $z_s=3$, respectively, and assuming the cosmological background models previously introduced in Sec. \ref{app}, the resulting deflection angle as well as the convergence are shown in the bottom panel of Figures \ref{fig3} (flat $\mu$HDM cosmology) and \ref{fig4} (flat minimal-matter cosmology), assuming $\theta=90^{\circ}$. Here the occurrence of negative $\kappa$-values simply reflects the fact that our model \eqref{eq:12} generates a local underdensity, $1+\delta(r)<1$, with the overall matter density $\rho$ being non-negative at any radius. Compared to the Newtonian case where $\kappa_{N}\lesssim 10^{-4}$, we again find that a face-on TeVeS filament may cause a significantly larger lensing signal, which is now on the order of $\kappa\sim\gamma\sim 10^{-3}$ within both TeVeS cosmologies. As the results of the $\mu$HDM and the minimal-matter cosmology approximately differ by a factor $1.5-2$ just as in \S \ref{app}, the order-of-magnitude lensing effects caused by TeVeS filaments are also in this case more or less cosmologically model-independent.

Close to the filament's axis, where $\kappa\sim 4\times 10^{-3}$, one can actually have a lensing signal $\kappa =\gamma =0.01$ assuming that the inclination angle is small, $\theta\lesssim 20^{\circ}$. Although such angles correspond to rather special configurations, we may conclude that also for our simple oscillation model, single TeVeS filaments {\it can} generate a lensing signal $\sim 0.01$, which is similar to our result in \S \ref{app}. However, note that the above discussion is based on the choice of equation \eqref{eq:12} and $\delta_{0}=4$. Considering a higher overdensity along its axis, even a face-on filament described by a similar fluctuating profile could easily create a shear field $\gamma\sim 0.01$ for $y\lesssim R_{f}$.

\section{Superimposing Filaments with Other Objects}
\label{superpos}

\begin{figure*}
\centering
\begin{minipage}[t]{8.5 cm}
\begin{center} 
\includegraphics[trim=30 0 0 0,width=0.9\textwidth]{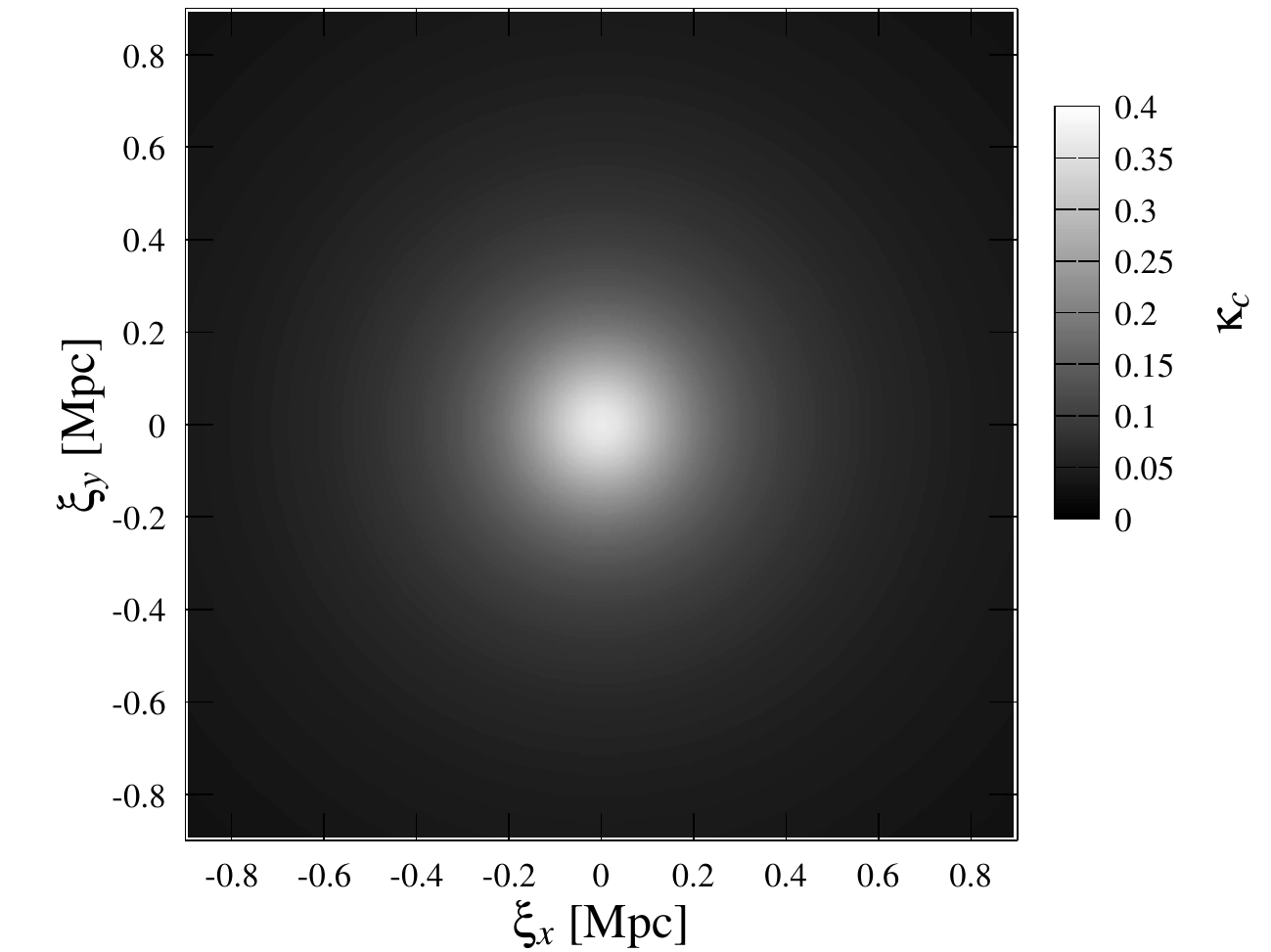}\\[0.3cm]
\includegraphics[trim=30 0 0 0,width=0.9\textwidth]{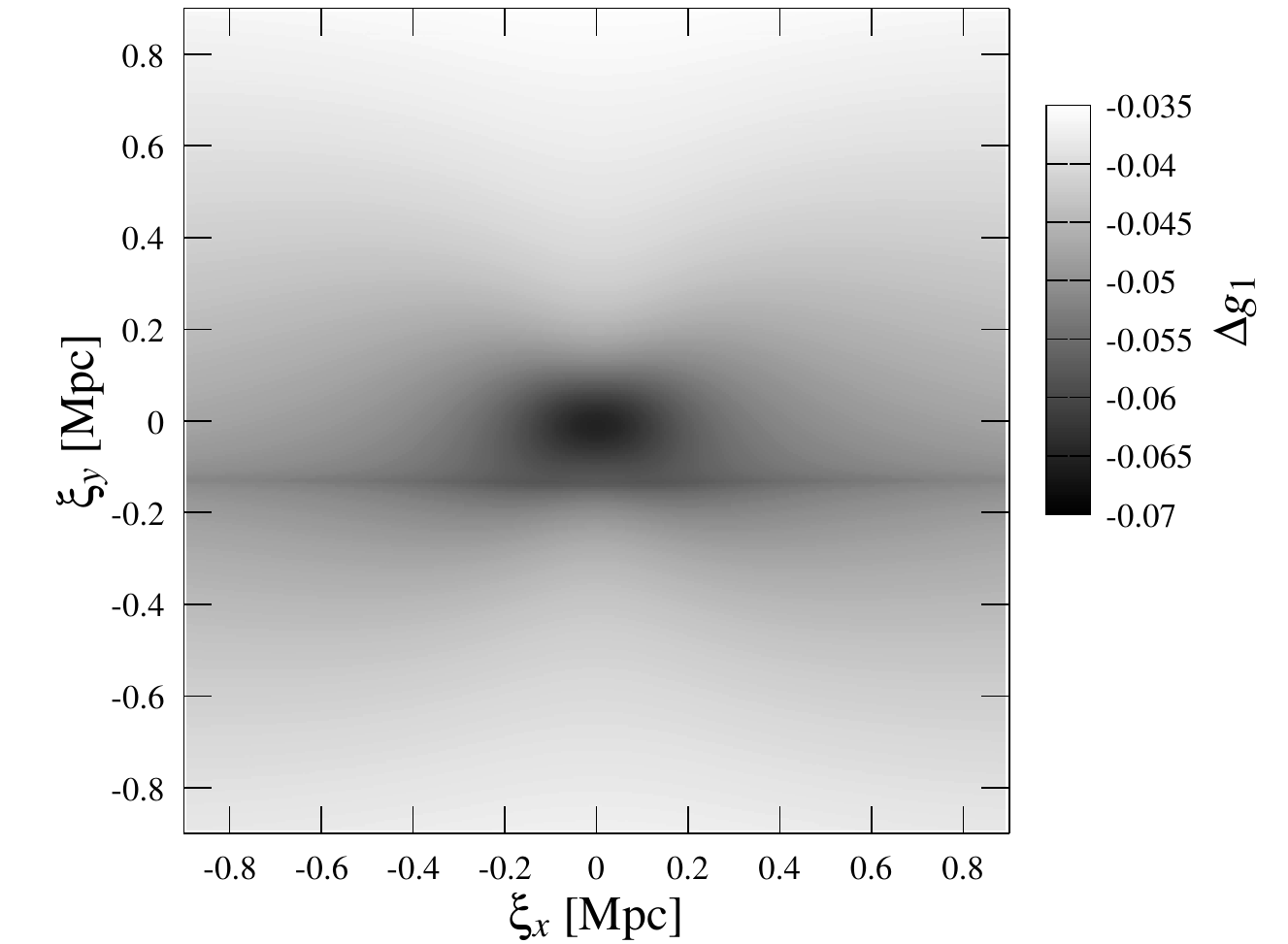}
\end{center}
  \end{minipage}
\qquad
 \begin{minipage}[t]{8.5 cm}
\begin{center}
\includegraphics[trim=30 0 0 0,width=0.9\textwidth]{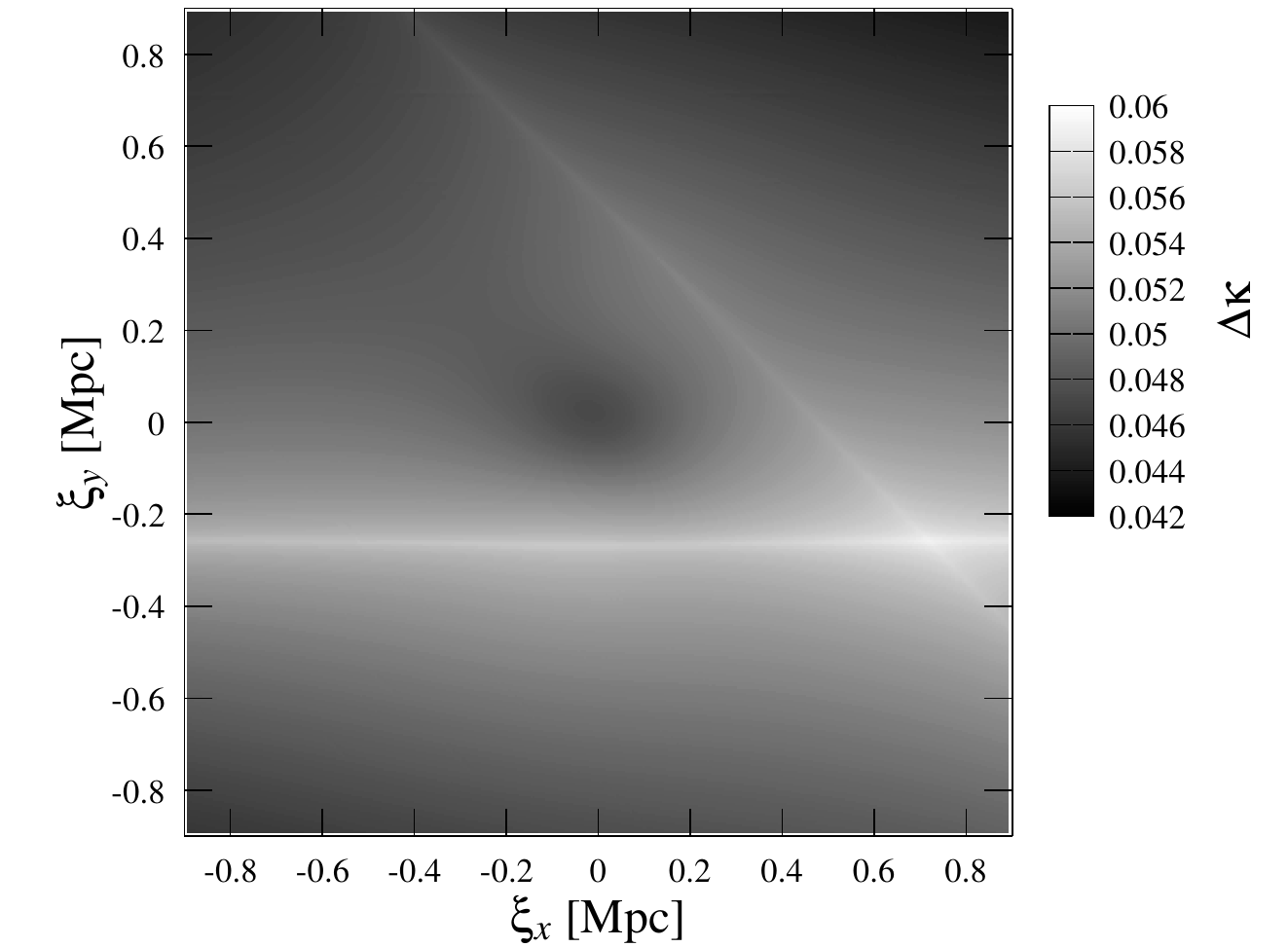}\\[0.3cm]
\includegraphics[trim=30 0 0 0,width=0.9\textwidth]{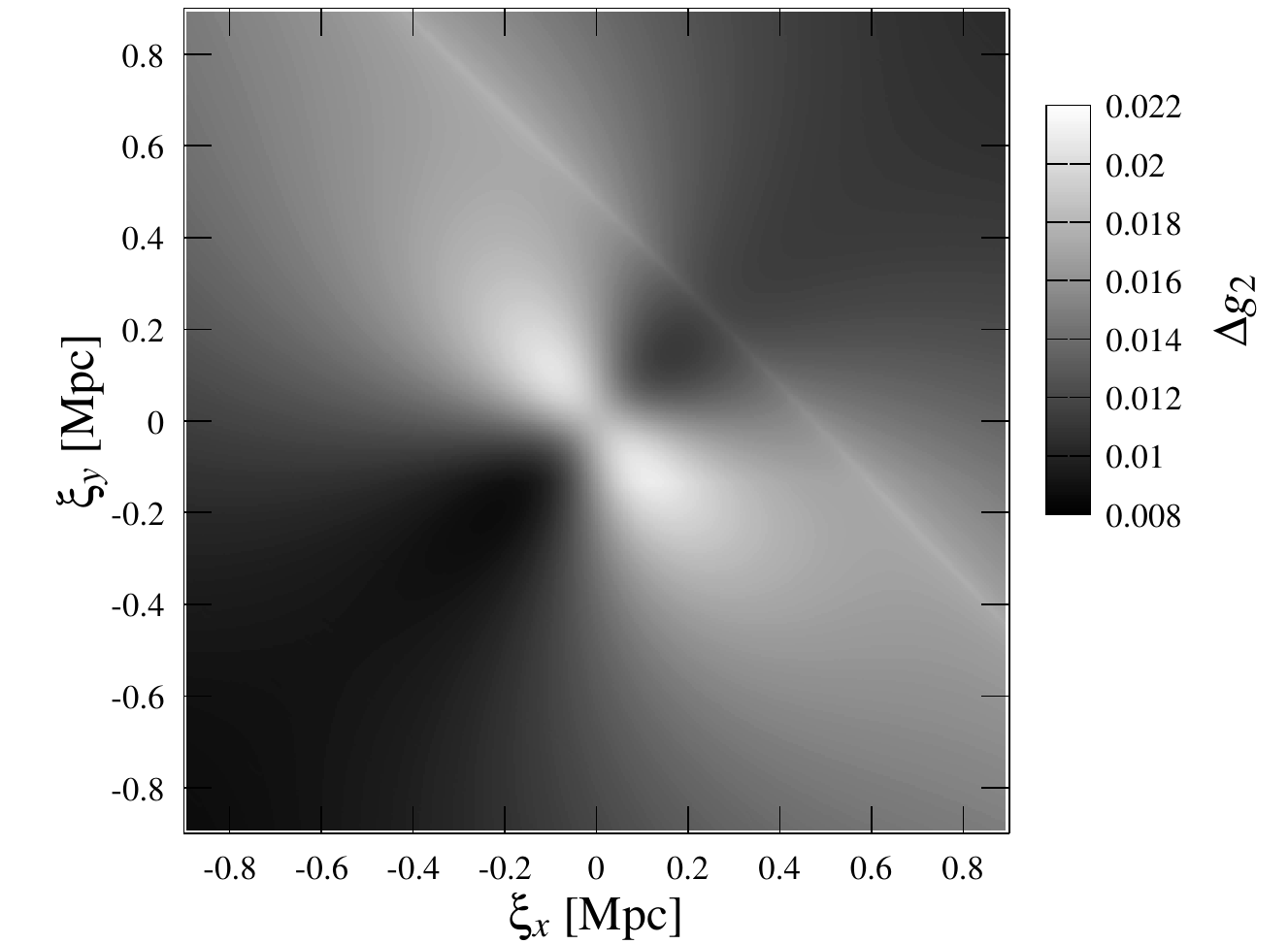}
\end{center}
\end{minipage}
\caption{Superposition of two filaments with a toy cluster along the line of sight; shown are the cluster's convergence map $\kappa_{c}$ in absence of any filamentary structures along the line of sight ({\it top left}) and the filaments' contribution $\Delta\kappa=\kappa_{tot}-\kappa_{c}$ to the total convergence ({\it top right}), as well as to the components of the reduced shear, $\Delta g_{1}=\gamma_{tot,1}/(1-\kappa_{tot})-\gamma_{c,1}/(1-\kappa_{c})$ and $\Delta g_{2}=\gamma_{tot,2}/(1-\kappa_{tot})-\gamma_{c,2}/(1-\kappa_{c})$, respectively ({\it bottom panel}).}
\label{fig5}
\end{figure*}

\begin{table}
\begin{center}
\caption{Parameters of the superimposed filaments in \S \ref{superpos} \label{table1}}
\begin{tabular}{cccccc}
\tableline
\tableline
\noalign{\smallskip}
 & P.A. & Incl.\footnotemark\footnotetext[1]{Inclination of the filament's axis to the line of sight} & Shift from Origin\footnotemark\footnotetext[2]{Shift of the filament's projection in the corresponding redshift plane} &  \\
Plane & (deg) & (deg) & (kpc) & Redshift $z$\\
\tableline
\noalign{\smallskip}
2 & 90 & 12 & (0,-150) & 0.25\\
3 & 45 & 45 & (600,0) & 0.30\\
\tableline
\end{tabular}
\end{center}
\end{table}

To demonstrate the contribution of filamentary structures to the lensing map of other objects, e.g. galaxy clusters, we superimpose two differently orientated filaments with a toy cluster along the line of sight, assuming the previously introduced $\mu$HDM cosmology and different redshifts for each component. If all objects are sufficiently far away from each other ($\gtrsim 100$Mpc), we may approximately treat them as isolated lenses at a certain redshift slice, i.e. the corresponding deflection angles can be calculated separately\footnotemark\footnotetext[8]{Note that, in general, one would have to solve the full non-linear TeVeS scalar field equation, which is beyond the scope of this work.}. Thus, we may resort to the well-known multiplane lens equation \citep{blandford,gl}:
\begin{equation}
\vec\eta = \frac{d_{s}}{d_{1}}\vec\xi_{1}-\sum_{i=1}^{n}d_{is}\vec{\hat\alpha}_{i}(\vec\xi_{i}),
\label{eq:15}
\end{equation}
where $n$ is the number of lens planes, $d_{ij}$ corresponds to the angular diameter distance between the $i$-th and the $j$-th plane, and $\vec\xi_{i}$ is recursively given by
\begin{equation}
\vec\xi_{i} = \frac{d_{i}}{d_{1}}\vec\xi_{1}-\sum_{j=1}^{i-1}d_{ji}\vec{\hat\alpha}_{j}(\vec\xi_{j}),\quad 2\leq i \leq n.
\label{eq:16}
\end{equation}
Comparing equation \eqref{eq:15} to the lens equation for a single lens plane, we identify the total deflection angle,
\begin{equation}
\vec{\hat\alpha}_{tot}(\vec\xi_{1}) = \vec{\hat\alpha}_{1}(\vec\xi_{1})+\sum_{i=2}^{n}\frac{d_{is}}{d_{1s}}\vec{\hat\alpha}_{i}(\vec\xi_{i}) =\vec{\hat\alpha}_{c}+\vec{\hat\alpha}_{f},
\label{eq:17}
\end{equation}
where $\vec{\hat\alpha}_{c}$ and $\vec{\hat\alpha}_{f}$ are the deflection angle of an isolated cluster at $z_{1}$ and an additional contribution due to the superimposed filaments, respectively. Analog to the case of a single plane, further lensing quantities such as the total convergence and the total shear can be calculated from equation \eqref{eq:17}, using the general relations introduced in \S \ref{model}. For simplicity, we shall assume that the cluster's TeVeS potential follows the ``quasi-isothermal" profile of \cite{tevesfit}:
\begin{equation}
\Phi(\vec r) = v^{2}\log\sqrt{1+\frac{|\vec r-\vec r_{0}|^{2}}{p^{2}}},
\label{eq:18}
\end{equation} 
where $v$ is the asymptotic circular velocity, $p$ is a scale length, and $\vec{r}_{0}$ is the center's position.

Concerning the numerical setup, we set $v^2=2\times 10^{6}$ (km s$^{-1})^2$ and $p=200$ kpc, fixing the cluster's redshift to $z_{1}=0.2$. Furthermore, we choose the uniform filament model discussed in \S \ref{model} and assume that filaments have a constant overdensity of $\delta = 20$ as well as the same characteristic radius $R_{f}=2.5$ $h^{-1}$ Mpc. While the cluster is centered at the origin ($\xi_{x}=\xi_{y}=0$), the two filaments are set up according to the parameters given in Table \ref{table1}. Finally, we place the source plane at a redshift of $z_{s}=1$. Note that this specific
setting corresponds to a more realistic lensing configuration compared to our analysis in the sections above, with our choice again being motivated by results based on a $\Lambda$CDM universe. 

From the top right panel of Figure \ref{fig5}, we see that the filaments' contribution to the total convergence map,
$\Delta\kappa=\kappa_{tot}-\kappa_{c}$ ($\kappa_{c}$ is the cluster's convergence map in absence of any filamentary structures along the line of
sight) is comparable to our previous findings, with the signal again being on the order of $0.01$. Also, note the distortion effects caused by
the cluster and the peak close to the region where the two filaments overlap. Obviously, the contribution pattern depends on the actual configuration
as well as on the type and amount of the considered objects along the line of sight, and can generally be quite complex. In addition, we present
the changes in the reduced shear components, $\Delta g_{1}=\gamma_{tot,1}/(1-\kappa_{tot})-\gamma_{c,1}/(1-\kappa_{c})$ and
$\Delta g_{2}=\gamma_{tot,2}/(1-\kappa_{tot})-\gamma_{c,2}/(1-\kappa_{c})$, due to the filaments' presence in the bottom panel of Figure \ref{fig5}.

At this point, we should emphasize that we have considered the impact of filamentary structures alone. Depending on their particular position along
the line of sight, additional (foreground) objects such as galaxies, galaxy clusters and/or voids might (locally) contribute on a comparable level or even
exceed the signal caused by filaments. Of course, this further complicates the interpretation of the corresponding lens mapping, and we conclude that, in general, extracting the filaments' contribution can pose quite a challenge.

\section{Conclusions}
\label{discussion}
In this work, we have analyzed the gravitational lensing effect by filamentary structures in TeVeS, a relativistic
formulation of the MOND paradigm.

For this purpose, we have set up two different cosmological models in TeVeS: the so-called $\mu$HDM cosmology including massive neutrinos on the order of $2$ eV which have already been proposed as a remedy for the discrepancies between dynamical and visible mass on cluster scales \citep{neutrinos2} as well as for the CMB
\citep{tevesneutrinocosmo}, and a simple minimal-matter cosmology accounting for a universe which is made up of baryons alone. Encouraged by several HDM simulations and the fact that filamentary structures are generic, we have assumed that the properties of such structures, i.e. their shape and relative densities, are similar in CDM and MOND/TeVeS scenarios independent of the particularly used cosmological background.

Modeling these filaments as infinite uniform mass cylinders, we have derived analytic expressions for their lensing properties in MOND/TeVeS and Newtonian/GR gravity. Regardless of the actual used cosmological background, we have shown that TeVeS filaments can account for quite a substantial contribution to the weak lensing convergence and shear field, $\kappa \sim \gamma \sim 0.01$, as well as to the amplification bias, $A^{-1}\sim 1.02$, which is even true outside but close $(y\sim 2R_{f})$ to the projected ``edges" of the filament's matter density. Exploring a simple oscillating density model of a filament and its surrounding area, we have found that the lensing signal in this case is generally smaller, but can still be of the same order, taking into account that the filamentary structures may be inclined to the line of sight by rather small angles ($\theta\lesssim 20^{\circ}$).
In addition, we find that there is fundamental difference between GR and MOND/TeVeS considering idealized cylindrically symmetric lens geometries: wherever the projected matter density is zero, there will be no distortion as well as no amplification effects, i.e. image and source will look exactly the same. In the context of MOND/TeVeS, however, this changes as one can have such effects in these regions.
Finally, we have demonstrated the impact of filaments onto the convergence map of other objects by considering superposition with a toy cluster along the line of sight. Again, our results have shown an additional contribution comparable to that of a single isolated filament and that the contribution pattern of filaments can be generally quite complex.

Here we have considered the lensing signal generated by single filaments alone. Simulating the cosmic web in a standard $\Lambda$CDM cosmology, \cite{dolag} have found a shear signal $\gamma\sim 0.01-0.02$ along filamentary structures, which seems quite similar to what MOND/TeVeS can do. Note, however, that this signal is entirely dominated by the simulation's galaxy clusters, with the filament's signal being much smaller, approximately on the order of $10^{-4}-10^{-3}$.

Although our analysis is mainly of theoretical interest, the above result points to an interesting possibility concerning recent measurements of weak lensing shear maps. For instance, the weak shear signal in the ``dark matter peak" of Abell 520 \citep{abell520} is roughly at a level of $0.02$, 
which is comparable to what filaments could produce in MOND/TeVeS, but not in Newtonian gravity (also cf. \cite{wedding}).
Therefore, we conclude that filamentary structures might actually be able to cause such anomalous lensing signals within the framework of MOND/TeVeS. 

In principle, the predicted difference in the weak lensing signal could also be used to test the validity of modified gravity. As several attempts to detect filaments by means of weak lensing methods have failed so far, e.g. the analysis of Abell $220$ and $223$ by \cite{dietrich}, this might already be a first hint to possible problems within MOND/TeVeS gravity. On the other hand, shear signals around $\gamma\sim 0.01$ are still rather small to be certainly detected by today's weak lensing observations, and lacking $N$-body structure formation simulations in the framework of MOND/TeVeS, we cannot even be sure about how filaments form and how they look like in a MONDian universe compared to the CDM case. Clearly, more investigation is needed to gain a better understanding about the impact of filamentary structures.
\section*{Acknowledgments}
We thank Johan Fynbo, Jens Hjorth and other members of the Dark Cosmology Centre
for very stimulating discussions, special thanks go to Steen Hansen for
organizing a Dark Matter Workshop in Copenhagen which inspired us to write this paper.
D.X. thanks Kristian Pedersen and Yi-Peng Jiang for helpful discussion.
M.F. thanks Matthias Bartelmann and Cosimo Fedeli for valuable comments on the manuscript.
We also thank the anonymous referee for useful suggestions.
B.F., M.F. and H.S.Z. acknowledge hospitality at the Dark Cosmology Centre.
B.F. is a research associate of the FNRS,
H.S.Z. acknowledges partial support from the National Natural Science Foundation of China (NSFC; Grant No. 10428308)
and a UK PPARC Advanced Fellowship.  
M.F. is supported by a scholarship from the Scottish Universities Physics Alliance (SUPA).
The Dark Cosmology Centre is funded by the Danish National Research Foundation.


\bibliographystyle{apj}
\bibliography{ref}

\end{document}